\documentclass[aps,prd,preprint,tightenlines,superscriptaddress,
amsfonts,amssymb,amsmath,byrevtex,showpacs,nofootinbib]{revtex4-2}
\usepackage[polish,english]{babel}
\usepackage[utf8]{inputenc}
\usepackage[T1]{fontenc}
\usepackage{ulem}
\usepackage{latexsym}
\usepackage{graphicx}
\usepackage{geometry}
\usepackage{epstopdf}
\usepackage{color}

\graphicspath{ {./}{./rys/} } 
\usepackage{xcolor}
\usepackage{pdfsync}
\usepackage{fancyhdr}
\usepackage{makeidx}
\usepackage[breaklinks,colorlinks,backref]{hyperref}
\hypersetup{
    colorlinks, 
    linktoc=all, 
    linkcolor=red,  
  citecolor=hyptxt,
  urlcolor=blue}
  \hypersetup{
  citebordercolor=,
  filebordercolor=green,
  linkbordercolor=blue
}
\definecolor{hyptxt}{rgb}{0.7, 0.4, 0.9}
\usepackage{bibtopic}
\usepackage{amsmath}

\newcommand{\RNumb}{\mathbb{R}}

\newcommand{\UnitOp}{\hat{1\kern-4.75pt 1}} 
\newcommand{\MatUnit}{1\kern-4.75pt 1} 
\newcommand{\Bra}[1]{\langle #1 \vert} 
\newcommand{\Ket}[1]{\vert #1 \rangle} 
\newcommand{\BraKet}[2]{\langle #1 \vert #2 \rangle} 
\newcommand{\AAver}[1]{\langle\kern-2pt\langle #1 \rangle\kern-2pt\rangle} 

\newcommand{\Amp}{\mathrm{Amp}} 
\newcommand{\Norm}[1]{\|\kern.3ex#1\kern.3ex \|} 
\newcommand{\Var}[1]{\mathrm{var}(#1)} 
\newcommand{\Group}[1]{\textrm{#1}} 

\newcommand{\DCosets}[3]{#1 \backslash #2/ #3}
\newcommand{\HW}[1]{\mathcal{HW}(#1)} 
\newcommand{\LHW}{\mathcal{LHW}(4)} 
\newcommand{\StateSpace}[1]{\mathcal{#1}} 
\newcommand{\ConfSpace}[1]{\textbf{#1}} 
\newcommand{\SpaceTime}[1]{\mathcal{#1}} 
\makeatletter
\newcommand{\DirectInt}{\mathop{\mathpalette\direct@int\relax}\!\int}
\newcommand{\direct@int}[2]{%
  \sbox\z@{%
    \m@th
    \ooalign{%
      \hidewidth$\demote@style{#1}{\bigoplus}$\hidewidth\cr
      $#1\phantom{\int}$\cr
    }%
  }%
  \wd\z@=\z@\box\z@
}
\newcommand{\demote@style}[2]{%
  \ifx#1\displaystyle\scriptstyle#2\else
    \raise.5\fontdimen22\textfont2\hbox{$\scriptscriptstyle#2$}%
  \fi
}
\makeatother
\newcommand{\Komentarz}[1]{}

\usepackage{accents}

\begin{document}

\title{Semiclassical causal geodesics: Schwarzschild spacetime case}


\author{Aleksandra P\c{e}drak} \email{aleksandra.pedrak@ncbj.gov.pl}
\affiliation{Department of Fundamental Research, National Centre for Nuclear
  Research, Pasteura 7, 02-093 Warszawa, Poland}

\author{Andrzej G\'{o}\'{z}d\'{z}}
\email{andrzej.gozdz@umcs.lublin.pl}
\affiliation{Institute of Physics, Maria Curie-Sk{\l}odowska
University, pl.  Marii Curie-Sk{\l}odowskiej 1, 20-031 Lublin, Poland}

\author{Patryk Mach} \email{patryk.mach@uj.edu.pl} \affiliation{Institute of
  Theoretical Physics, Jagiellonian University in Krak\'{o}w,
  ul. prof. S. Łojasiewicza 11, 30-348 Kraków, Poland}

\author{W{\l}odzimierz Piechocki} \email{wlodzimierz.piechocki@ncbj.gov.pl}
\affiliation{Department of Fundamental Research, National Centre for Nuclear
  Research, Pasteura 7, 02-093 Warszawa, Poland}

\date{\today}

\begin{abstract}
We propose an extension of the integral quantization, applied previously to the
semiclassical description of a test particle in the flat Minkowski spacetime, to the Schwarzschild spacetime. This extension yields a local integral quantization. We calculate the transition amplitudes between coherent states of a particle in a mass layer, which effectively leads to a Klein-Gordon-type dynamics. The calculated probability distributions reproduce locally the classical radial geodesic motion in the Schwarzschild spacetime. The local integral quantization presented here seems to be applicable to semiclassical quantization of geodesic motion in other curved spacetimes.
\end{abstract}


\maketitle

\tableofcontents

\section{Introduction}\label{Int}

General relativity (GR) proves to be successful in describing most of known
classical gravitational systems, especially with low curvature of
spacetime. Quantum theory (QT) describing, for instance, the world of elementary
particles in flat spacetime is successful as well.  It is believed that near
gravitational singularities, where GR breaks down, quantum effects should be
taken into account, and both theories GR and QT should be unified into a theory
of quantum gravity (QG). However, this idea turns out to be extremely difficult to realize.  The struggle for the construction of the theory of quantum
gravity has lasted for more than fifty years.  The hope that string theory (see,
e.g., \cite{Green}), loop quantum gravity (see, e.g., \cite{Gam,Thi,Rov}),
causal dynamical triangulations \cite{cdt21,cdt12}, integral quantization (to be
applied in the present paper), or other theory could play that role has not
been fulfilled yet. The main reason seems to be related to the lack of experimental data
concerning extremal gravitational fields that could be used to reduce the freedom in
choosing the mathematical structures underlying those theories.

We believe that quantum effects may play a role in the description of particles plunging
into the black hole, a process typically associated, for example, with the black hole
information paradox. In this paper, we propose a semiclassical  quantization of
the motion of a spinless test particle on the fixed Schwarzschild background metric,
based on the coherent states method (see, \cite{Per,Klauder,Proc} and references therein),
and developed for the case of a free particle in the flat Minkowski spacetime
in \cite{paper1,paper2}. For simplicity, we mainly focus on the radial motion.
We also believe that investigating various quantization methods of a (classically geodesic)
motion in curved spacetimes can shed some light on the possible dynamics of particles
in the vicinity of black holes, as well as on the freedom available in the quantum description.

In this paper, we introduced the concept of local integral quantization (LIQ).
In this method the quantization is performed on the cotangent bundle of the spacetime manifold,
where the position on the manifold is a parameter and not a quantized quantity.
The quantization procedure is based on a properly selected system of coherent states.
Our construction of the coherent states is based on the Lorentz--Heisenberg--Weyl group, which we also introduce in this paper. To test the LIQ method, we computed the quantum transition
probabilities for a Klein–Gordon-type particle and compared the expectation values
for the particle position at a fixed time with the classical radial geodesics.

The paper is organized as follows: In Sec.~\ref{Time} we present radial
geodesics. The integral quantization in curved spacetime is presented in
Sec.~\ref{Integral}. The coherent states based on the representation of the Lorentz--Heisenberg--Weyl group are defined in Sec.~\ref{LLL}. Specific model is presented
in Sec.~\ref{Spec}. Section \ref{Sem} is devoted to the semiclassical motion of
a test particle. We conclude in Sec.~\ref{Con}.

In numerical calculations we apply the Planck units by setting $c = 1 = G = \hbar$,
where $c$ is the speed of light, $G$ is the gravitational constant, and $\hbar$
denotes the Planck constant. This choice of units renders the set
$\{p_\mu,x^\mu,m\}$ dimensionless.

\section{Timelike and null geodesics}\label{Time}

\subsection{Geodesic equations}

Schwarzschild geodesics have been studied extensively in many works, starting from \cite{droste,hagihara}. A recent approach based on the Biermann-Weierstrass formula can be found in \cite{CM2022,CM2023}.

In the standard Schwarzschild coordinates $(\bar t, r, \theta, \varphi)$, the
Schwarzschild metric can be written as
\begin{equation}
\label{droste}
g = -N d \bar t^2 + N^{-1}dr^2 + r^2 d \theta^2 + r^2 \sin^2 \theta d \varphi^2,
\end{equation}
where $N = 1 - 2M/r$ and $M > 0$ is the Schwarzschild mass. A system of coordinates $(t,r,\theta,\varphi)$ regular at the horizon $r = 2M$ can be introduced by
a suitable change of the time coordinate
\begin{equation}
dt = d\bar t + \left[ \frac{1}{N(r)} - \eta(r) \right] dr,
\end{equation}
where $\eta(r)$ is a given function of the areal radius $r$. This transforms the
Schwarzschild metric to the form
\begin{equation}
\label{schwbl}
g = -N dt^2 + 2 (1 - N \eta) dt dr + \eta ( 2 - N \eta) dr^2 + r^2 d \theta^2
+ r^2 \sin^2 \theta d \varphi^2.
\end{equation}
A frequent choice $\eta \equiv 1$ yields the so-called ingoing Eddington-Finkelstein
coordinates, regular at $r = 2M$. In this case, the metric reads
\begin{equation}
\label{schwef}
g = - \left( 1 - \frac{2M}{r} \right) dt^2 + \frac{4M}{r} dt dr + \left( 1 + \frac{2M}{r} \right) dr^2 + r^2 d \theta^2
+ r^2 \sin^2 \theta d \varphi^2.
\end{equation}
Setting $\eta = 1/N$, we recover Eq.\ (\ref{droste}). In the following, we will work in the ingoing Eddington-Finkelstein coordinates with $\eta \equiv 1$.

Geodesic equations can be written in a Hamiltonian form, which to some extent
motivates  our quantization framework. Defining
\begin{equation}
\label{gHam}
 H = \frac{1}{2} g^{\mu\nu}(x^\alpha) p_{\mu} p_{\nu},
\end{equation}
one can write geodesic equations as
\begin{equation}
\label{hamilton_eqs}
\frac{dx^\mu}{d\tilde{s}} = \frac{\partial H}{\partial p_\mu},
\quad \frac{dp_\mu}{d\tilde{s}} = - \frac{\partial H}{\partial x^\mu}.
\end{equation}
Here the four momentum is defined as $p^\mu = d x^\mu/d\tilde{s}$. It is
normalized so that
$H = \frac{1}{2} g^{\mu \nu} p_\mu p_\nu = - \frac{1}{2} m^2$, where $m \ge 0$
is the particle rest mass. For $m > 0$, this implies that the affine parameter $\tilde s$ is
related to the proper time $\tilde \tau$ by $\tilde s = \tilde \tau/m$.

Assuming the Schwarzschild metric of the form (\ref{schwbl}), one can write the
Hamiltonian $H$ as
\begin{equation}
H = \frac{1}{2} \left[ g^{tt}(r)p_t^2 + 2 g^{tr}(r) p_t p_r + g^{rr}(r) p_r^2
+ \frac{1}{r^2} \left( p_\theta^2
+ \frac{p_\varphi^2}{\sin^2 \theta} \right) \right].
\end{equation}
The energy $E = -p_t$, the azimuthal component of the
angular momentum $l_z = p_\varphi$, the square of the total angular momentum
$l^2 = p_\theta^2 + p_\varphi^2/\sin^2 \theta$, and the Hamiltonian
$H = - \frac{1}{2} m^2$ are conserved. This allows us to write geodesic equations
as
\begin{subequations}
\label{eqsofmotion1}
\begin{alignat}{1}
\frac{dr}{d\tilde{s}} &= \epsilon_r \sqrt{E^2 - \tilde U_{l,m}(r)},\\
\frac{d\varphi}{d\tilde{s}} &=  \frac{l_{z}}{r^{2}\sin^2 \theta}, \\
\frac{d\theta}{d\tilde{s}} &= \frac{\epsilon_\theta}{r^{2}}
\sqrt{ l^{2} -\frac{l_{z}^2}{\sin^2 \theta}},\\
\frac{dt}{d\tilde{s}}  &=  \frac{E r}{r - 2 M} + \epsilon_r \frac{2 M}{r - 2 M}\sqrt{E^2 - \tilde U_{l,m}(r)}. \label{dtdstilde}
\end{alignat}
\end{subequations}
where
\begin{equation}
\tilde U_{l,m}(r) = \left(1 - \frac{2M}{r} \right)
\left( \delta_m + \frac{l^2}{r^2} \right)
\end{equation}
is the radial effective potential. The constant $\delta_m$ is defined as
\begin{equation}
\delta_m = \begin{cases}
m^2 & \text{for timelike geodesics,} \\
0 & \text{for null geodesics,}
\end{cases}
\end{equation}
and the signs $\epsilon_\theta = \pm 1$ and $\epsilon_r = \pm 1$, correspond to
the directions of motion.

There are a few equivalent ways to partially decouple the above set of
equations. The easiest (but least general) is to choose a coordinate system so
that the motion occurs in the equatorial plane. Alternatively, one can introduce
a new variable, the so-called true anomaly, according to equations given for
instance in \cite{kostic_2012}. The third option is to reparametrize geodesics
by introducing a new parameter $\bar s$ defined by
\begin{equation}
\frac{dx^\mu}{d \bar s} = r^2 \frac{dx^\mu}{d \tilde s} \; \;
\text{or} \; \; \tilde s = \int_0^{\bar s} r^2 ds.
\end{equation}
This is a direct counterpart of a reparametrization introduced by Mino in
\cite{mino_2003} in the context of the Kerr spacetime; consequently, we will
refer to the parameter $\bar s$ as the Mino time. With respect to the Mino time,
Schwarzschild geodesic equations take the form
\begin{subequations}
\label{eqsofmotion2}
\begin{eqnarray}
        \frac{dr}{d\bar{s}} & = & \epsilon_r r^2 \sqrt{E^2 - \tilde U_{l,m}(r)}, \label{minor}\\
    \frac{d\varphi}{d\bar{s}} & = & \frac{l_{z}}{\sin^2 \theta}, \\
    \frac{d\theta}{d\bar{s}} & = & \epsilon_\theta \sqrt{ l^{2} -\frac{l_{z}^2}{\sin^2 \theta}}, \label{minotheta}\\
    \frac{dt}{d\bar{s}} & = & \frac{E r^3}{r - 2 M} + \epsilon_r \frac{2 M r^2}{r - 2 M}\sqrt{E^2 - \tilde U_{l,m}(r)}.
\end{eqnarray}
\end{subequations}
In the following, we will work in dimensionless rescaled variables. For timelike geodesics with $m > 0$, we set
\begin{alignat}{7}
t = M T, &\; r = M \xi, &\; E  =  m \varepsilon, &\; l =  M m \lambda, &\;
l_z =
M m \lambda_z,
\end{alignat}
so that, in particular, $N = 1 - 2/\xi$. A new Mino time $s$ is defined by
\begin{equation}\label{aff}
\bar{s} = \frac{s}{M m}.
\end{equation}
For null geodesics, one can replace the parameter $m$ with an arbitrary positive constant $\mu$ with a dimension of the mass. In terms of dimensionless variables, geodesic equations take the form
\begin{subequations}
\label{eqsofmotion3}
\begin{eqnarray}
        \frac{d\xi}{ds} & = & \epsilon_r \xi^2 \sqrt{\varepsilon^2 -
          U_{\lambda}(\xi)}, \label{eqxi} \\
    \frac{d\varphi}{d s} & = & \frac{\lambda_{z}}{\sin^2 \theta}, \label{eqphi}
    \\
    \frac{d\theta}{ds} & = & \epsilon_\theta \sqrt{ \lambda^2
      -\frac{\lambda_z^2}{\sin^2 \theta}}, \label{eqtheta}\\
    \frac{dT}{ds} & = & \frac{\varepsilon \xi^3}{\xi - 2} + \epsilon_r \frac{2 \xi^2}{\xi - 2}\sqrt{\varepsilon^2 - U_{\lambda}(\xi)},\label{T_mot}
\end{eqnarray}
\end{subequations}
where
\begin{equation}
U_{\lambda}(\xi) = \left(1 - \frac{2}{\xi} \right) \left( \delta +
\frac{\lambda^2}{\xi^2} \right).
\end{equation}
Here
\begin{equation}
\delta = \begin{cases}
1 & \text{for timelike geodesics,} \\
0 & \text{for null geodesics.}
\end{cases}
\end{equation}

\subsection{Time orientation}
\label{sec:timeorientation}

We continue working in the ingoing Eddington--Finkelstein coordinates. A vector normal to hypersurfaces of constant time $t$ has the components $n_\mu = (-1,0,0,0)$, or alternatively
\begin{equation}
    n = \left( 1 + \frac{2M}{r} \right) \partial_t - \frac{2M}{r} \partial_r.
\end{equation}
It satisfies $g(n,n) = g^{tt} = - (1 + 2M/r)$, and hence it is always timelike. We consider this vector to be future-directed.

A future directed four-momentum associated with a time-like geodesic has to satisfy
\begin{equation}
    p(n) = -p^t = - \frac{1}{1 - \frac{2M}{r}} \left[ E + \epsilon_r \frac{2 M}{r} \sqrt{E^2 - \tilde U_{l,m}(r)} \right] < 0,
\end{equation}
but it is difficult to conclude about the sign of $p(n)$ from this form.

Observe that for $r > 2M$ the vector $X = \partial_t$ is also timelike, as $g(X,X) = g_{tt} = -(1 - 2M/r) < 0$. Since $g(n,X) = -1$, it is also future-directed. The condition
\begin{equation}
    p(X) = p_t = -E < 0
\end{equation}
implies that all momenta with $E > 0$ are future-directed for $r > 2M$.

For $r < 2M$, we take $Y = 2M \partial_t + (r - 2M) \partial_r$. This vector is timelike for $r < 2M$, since
$g(Y,Y) = r(r - 2M) < 0$, and it is also future-directed for $r < 2M$, as $g(n,Y) = -2 M < 0$. Now,
\begin{equation}
    p(Y) = 2 M p_t + (r - 2M )p_r = \epsilon_r r \sqrt{E^2 - \tilde U_{l,m}(r)}.
\end{equation}
Hence, for $r < 2M$ momenta with $\epsilon_r = - 1$ are future-directed, while those with $\epsilon_r = + 1$ are past-directed. Note that for $r < 2M$ one can have a future-directed timelike geodesic characterized by $E < 0$ and $\epsilon_r = -1$. On the other hand, such geodesics do not originate in the region $r > 2M$ (block I in the maximal Kruskal-Szekeres extension).

\subsection{Radial geodesics}

Radial geodesics are characterized by $\lambda = 0$ and $\lambda_z = 0$. In this case, Eqs.\ (\ref{eqsofmotion3}) reduce to
\begin{eqnarray}
\label{radialreq}
    \frac{d\xi}{ds} & = &  \epsilon_r \sqrt{(\varepsilon^2 - \delta) \xi^4 + 2 \delta \xi^3}, \\
\frac{dT}{ds} & = & \frac{1}{\xi - 2} \left[ \varepsilon \xi^3 + 2 \epsilon_r \sqrt{(\varepsilon^2 - \delta) \xi^4 + 2 \delta \xi^3}  \right].
    \label{radialTeq}
\end{eqnarray}
and $d \theta/ds = 0$, $d \varphi/ds = 0$.
Note that Eq.\ (\ref{radialTeq}) can also be written as
\begin{equation}
    \frac{dT}{ds} = \frac{\xi^3 \left[ 4 \delta + \varepsilon^2 \xi (2 + \xi) \right]}{\varepsilon \xi^3 - 2 \epsilon_r \sqrt{(\varepsilon^2 - \delta) \xi^4 + 2 \delta \xi^3}}.
\end{equation}
Let $\sigma = \epsilon_r \mathrm{sgn}(\varepsilon)$. For $\sigma = -1$, the right-hand side of this expression remains regular at the horizon $\xi = 2$. On the other hand, for $\sigma = +1$, it diverges for $\xi \to 2$.

The simplicity of the above equations allows us to parameterize the solutions with the time coordinate $T$. By combining Eqs.\ (\ref{radialreq}) and (\ref{radialTeq}), we get
\begin{equation}
    \frac{dT}{d\xi} = \frac{1}{\xi - 2} \left[ 2 + \epsilon_r \varepsilon \xi \sqrt{\frac{\xi}{(\varepsilon^2 - \delta) \xi + 2 \delta}} \right].
\end{equation}
Introducing
\begin{equation}
e=|\varepsilon|,
\qquad
q(\xi) = \sqrt{e^2 - \delta + \frac{2 \delta}{\xi}},
\end{equation}
we can write this equation as
\begin{equation}
\label{radialsigmaq}
    \frac{d T}{d\xi} = \frac{1}{\xi - 2} \left[ 2 + \frac{\sigma e \xi}{q(\xi)} \right],
\end{equation}
where, as before, $\sigma = \epsilon_r \mathrm{sgn}(\varepsilon)$.

For $\delta = 0$ (null geodesics) the above equation can be reduced to
\begin{equation}
    \frac{dT}{d\xi} = \frac{\sigma \xi + 2}{\xi - 2}.
\end{equation}
In this case, for $\sigma = -1$, one obtains $dT/d\xi = -1$, and thus $T = - \xi + C$, where $C$ denotes an integration constant. For $\sigma = +1$, we get
\begin{equation}
    T = \xi + 4 \ln |\xi - 2| + C.
\end{equation}

For timelike radial geodesics ($\delta=1$), the function $q(\xi)$ reads
\begin{equation}
    q(\xi) = \sqrt{e^2 - 1 + \frac{2}{\xi}}.
\end{equation}
To provide explicitly real solutions of Eq.\ (\ref{radialsigmaq}), it is convenient to introduce $a = 1 - e^2$. For $e\neq 1$, define the auxiliary function
\begin{equation}
F_e(\xi) = 2\ln\xi + 4\ln[ e + q(\xi)] + \frac{e \xi q(\xi)}{a} +\frac{2 e (1 + 2a)}{a} \mathcal A_a(q(\xi)),
\label{radialF}
\end{equation}
where
\begin{equation}
\mathcal A_a(q)=
\begin{cases}
\displaystyle
\frac{1}{\sqrt a}
\arctan\left(\frac{q}{\sqrt a}\right),
& a>0,\\[10pt]
\displaystyle
\frac{1}{2\sqrt{-a}}
\ln\left|
\frac{q-\sqrt{-a}}{q+\sqrt{-a}}
\right|,
& a<0.
\end{cases}
\end{equation}
The function $F_e$ satisfies
\begin{equation}
F_e'(\xi)=\frac{1}{\xi-2}
\left(2-\frac{e\xi}{q(\xi)}\right).
\end{equation}
Consequently, the solution of
Eq.~(\ref{radialsigmaq}), subject to the
initial condition $T(\xi_0)=T_0$, can be
written as
\begin{equation}
T(\xi) = T_0
-\sigma\left[F_e(\xi)-F_e(\xi_0)\right]
+ 2(1 + \sigma)
\ln \left| \frac{\xi - 2}{\xi_0 - 2} \right|.
\label{radialTsol}
\end{equation}
For the special case $e=1$, the function $F_e$
should be replaced by
\begin{equation}
F_1(\xi)=
4\ln\left(1+\sqrt{\frac{\xi}{2}}\right)
-4\sqrt{\frac{\xi}{2}}
-\frac{4}{3}\left(\frac{\xi}{2}\right)^{3/2}.
\label{radialF1}
\end{equation}
Equation~(\ref{radialTsol}) applies to both
signs of the conserved energy and both radial
directions, provided that $\varepsilon\neq 0$.
The two possible values of $\sigma$ distinguish
the branches that are regular ($\sigma=-1$)
or logarithmically divergent ($\sigma=+1$)
at $\xi=2$.

The case with $\delta = 1$ and $\varepsilon=0$ can be treated
separately. Equation (\ref{radialsigmaq}) then
reduces to
\begin{equation}
\frac{dT}{d\xi}=\frac{2}{\xi-2},
\end{equation}
and hence
\begin{equation}
T(\xi)=T_0+
2\ln\left|\frac{\xi-2}{\xi_0-2}\right|.
\label{radialTzero}
\end{equation}
Such timelike geodesics are confined to the
black hole interior, $0<\xi<2$. They are future-directed for
$\epsilon_r=-1$ and past-directed for
$\epsilon_r=+1$. In particular, for the
future-directed branch,
\begin{equation}
\frac{d\xi}{ds}
=-\xi^2\sqrt{\frac{2}{\xi}-1}<0,
\qquad
\frac{dT}{ds} = \frac{2\xi}{\sqrt{\frac{2}{\xi} - 1}} > 0.
\end{equation}

For $0<e<1$, radial timelike geodesics
have a turning point at
\begin{equation}
\xi_{\mathrm{max}}=\frac{2}{1-e^2}.
\label{radialturn}
\end{equation}
A geodesic passing through this point changes
the sign of $\epsilon_r$, while its conserved
energy remains unchanged. Consequently,
$\sigma$ also changes sign. The two branches
of Eq.~(\ref{radialTsol}) can be matched at
$\xi=\xi_{\mathrm{max}}$ by imposing the
same value $T_{\mathrm{max}}$ at this point. Denoting
$T_{\mathrm{max}}=T(\xi_{\mathrm{max}})$, we obtain
\begin{equation}
T(\xi)=
\begin{cases}
T_{\mathrm{max}}-F_e(\xi)
+F_e(\xi_{\mathrm{max}})
+4\ln\left(
\dfrac{\xi-2}{\xi_{\mathrm{max}}-2}
\right),
&\sigma=+1,\\[10pt]
T_{\mathrm{max}}+F_e(\xi)
-F_e(\xi_{\mathrm{max}}),
&\sigma=-1.
\end{cases}
\label{radialTturn}
\end{equation}
Although $dT/d\xi$ diverges at the turning
point, the geodesic remains smooth when
parametrized by the Mino parameter $s$.

As discussed in Sec.\ \ref{sec:timeorientation}, the sign
of the conserved energy does not uniquely
determine the time orientation of a geodesic
inside the black hole. In particular,
geodesics with $\varepsilon<0$ and
$\epsilon_r=-1$ are future-directed in
the region $0<\xi<2$. For these geodesics
$\sigma=+1$, and Eq.~(\ref{radialTsol})
implies
\begin{equation}
T(\xi)=4\ln(2-\xi)+O(1),
\qquad \xi\to2^-.
\end{equation}
Thus, when followed backwards in time,
these geodesics approach the horizon at
$T\to-\infty$, although the horizon is
reached in a finite proper time. In the
maximal Kruskal extension, they originate
in the second asymptotically flat region
(region III) and cross the left branch
of the future event horizon into the
black hole interior (region II).

\subsection{Cartesian coordinates}
\label{Cartesian}

In what follows, we will work in the system of Cartesian coordinates $(t,x,y,z)$, related to the ingoing Eddington-Finkelstein coordinates $(t,r,\theta,\varphi)$ by
\begin{subequations}
\label{cartesian}
\begin{eqnarray}
t & = & t, \\
x & = & r \sin \theta \cos \varphi, \\
y & = & r \sin \theta \sin \varphi, \\
z & = & r \cos \theta.
\end{eqnarray}
\end{subequations}
The coordinates $(t,x,y,z)$ can be understood as a counterpart of the Kerr-Schild system of coordinates for the Schwarzschild metric. In coordinates $(t,x,y,z)$ the Schwarzchild metric (\ref{schwbl}) takes the form
\begin{equation}
g_{\mu \nu} = \left(
\begin{array}{cccc}
-N & \frac{x (1 - N)}{r} & \frac{y (1 - N)}{r} & \frac{z (1 - N)}{r} \\
\frac{x (1 - N)}{r} & \frac{x^2 (2 - N)+y^2+z^2}{r^2} & \frac{x y \left( 1 - N \right)}{r^2} & \frac{x z ( 1 - N)}{r^2} \\
\frac{y (1 - N)}{r} & \frac{x y \left( 1 - N \right)}{r^2} & \frac{y^2 (2 - N) + x^2 + z^2}{r^2} & \frac{y z (1 - N)}{r^2} \\
\frac{z (1 - N)}{r} & \frac{x z (1 - N)}{r^2} & \frac{y z (1 - N)}{r^2} & \frac{ z^2 (2 - N) + x^2 + y^2}{r^2}
\end{array}
\right).
\end{equation}
The contravariant components read
\begin{equation}
g^{\mu \nu} = \left(
\begin{array}{cccc}
   (N -2) & \frac{x (1 - N)}{r} & \frac{y
   (1 - N)}{r} & \frac{z (1 - N)}{r} \\
 \frac{x (1 - N)}{r} & \frac{N x^2+y^2+z^2}{r^2} &
   \frac{(N - 1) x y}{r^2} & \frac{(N - 1) x z}{r^2} \\
 \frac{y (1 - N)}{r} & \frac{(N - 1) x y}{r^2} &
   \frac{N y^2+x^2+z^2}{r^2} & \frac{(N - 1) y z}{r^2} \\
 \frac{z (1 - N)}{r} & \frac{(N - 1) x z}{r^2} &
   \frac{(N - 1) y z}{r^2} & \frac{N z^2+x^2+y^2}{r^2} \\
\end{array}
\right).
\end{equation}

The so-called Kerr-Schild form the coordinates $(t,x,y,z)$ allows us to express the square of a covector $p$ with components $p_\mu = (p_t,p_x,p_y,p_z)$ as
\begin{equation}
g^{\mu \nu} p_\mu p_\nu = - p_t^2 + p_x^2 + p_y^2 + p_z^2 - \frac{2M (-p_t r + p_x x + p_y y + p_z z)^2}{r^3},
\end{equation}
where the term $- p_t^2 + p_x^2 + p_y^2 + p_z^2$ corresponds to a flat Minkowski spacetime.

\begin{figure}
\includegraphics[width=0.5\textwidth]{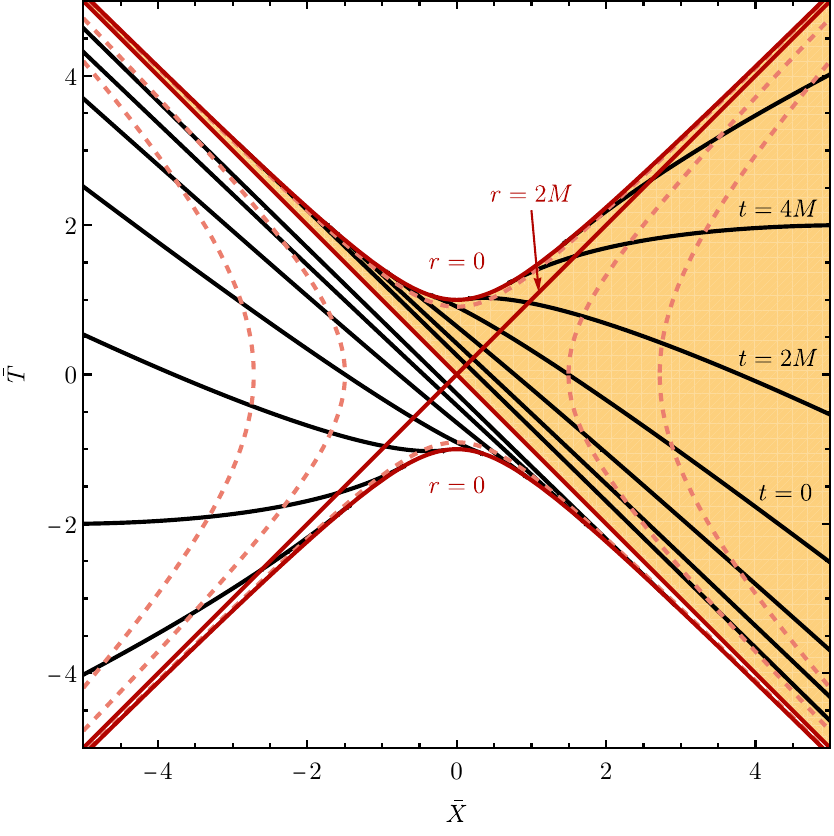}
\caption{\label{figkruskal} Hypersurfaces of constant Eddington--Finkelstein time $t$ on the Kruskal diagram. The Schwarzschild region is marked in orange. Solid black lines depict surfaces of constant time $t$. Surfaces of constant $r$ are plotted with red lines. Solid red lines depict the surfaces with $r = 0$ and $r = 2M$ (horizon).}
\end{figure}

\begin{figure}
\includegraphics[width=\textwidth]{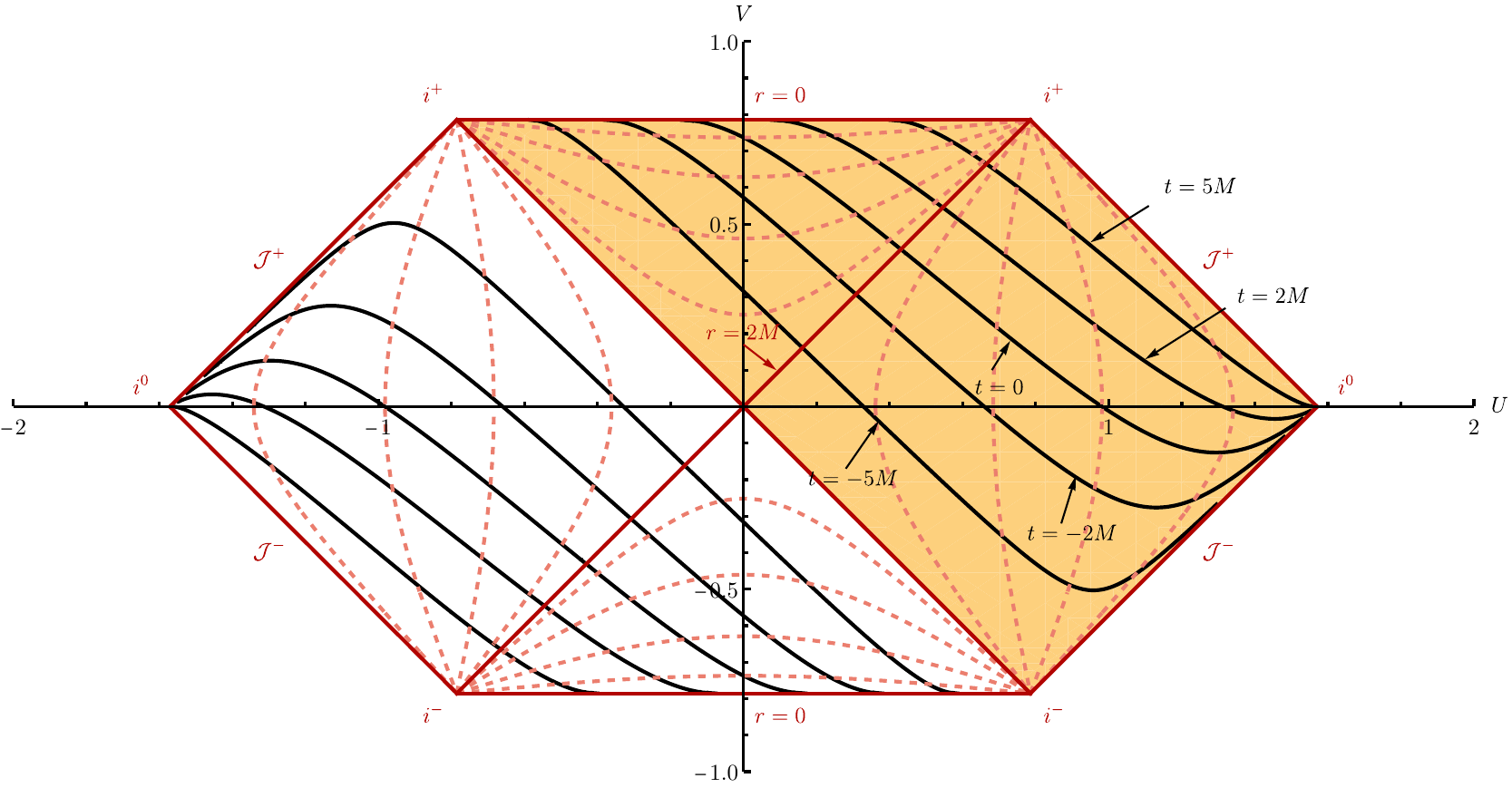}
\caption{\label{figpenrose}Hypersurfaces of constant Eddington--Finkelstein time on the Penrose--Carter diagram. We use the same color-scheme as in Fig.\ \ref{figkruskal}. The Schwarzschild region is marked in orange. Solid black lines depict surfaces of constant time $t$. Surfaces of constant $r$ are plotted with red lines. Solid red lines depict the surfaces with $r = 0$ and $r = 2M$ (horizon).}
\end{figure}

In the maximal analytic Kruskal-Szekeres extension of the Schwarzschild spacetime, the range $(t,x,y,z) \in \mathbb R^4$ covers regions I and II, marked in orange in Fig.\ \ref{figkruskal}. In Figure \ref{figpenrose} we provide the corresponding Penrose-Carter diagram for the same maximal extension. In both plots, thick black lines represent hypersurfaces of constant time $t$. The details of the construction of both diagrams are standard, and hence we decided to omit them in this paper. We would only like to emphasize that in both figures, the graphs representing the hypersurfaces of constant time $t$ are obtained as a result of numerical calculation, i.e., they are not merely schematic drawings.

\section{Integral quantization in curved spacetime}
\label{Integral}

In this paper we use the integral quantization (IQ) method generalized
to the case of curved spacetime. For a comprehensive presentation of this method
applied to the case of flat spacetime we recommend our recent article
\cite{paper1} and its direct application to the description of semiclassical
geodesics in Minkowski spacetime \cite{paper2}.

In both cases, we have used the IQ method based on the Heisenberg-Weyl group. For
an application of the IQ method based on the affine group, see our papers in
astrophysics \cite{ast1,ast2,ast3,ast4} and cosmology \cite{cos1,cos2,
  cos3,met}.

The general idea of the IQ method requires introducing a locally compact group
$\Group{G}_Q$, called the quantization group. Generalization of this method to
a local integral quantization (LIQ) suggests introduction of a larger group
$\Group{G}_0 \supset \Group{G}_Q$ named the structure group. This group has to
contain the full set of the required classical (rather semi-classical) and quantum
variables describing the physical system under consideration. This makes the
formalism more flexible. This also means that one can establish a mapping from the structure group $\Group{G}_0$
onto the classical configuration space $\ConfSpace{X}$ of the system. On the
other hand, the quantized part of the classical configuration space has to remain in
one-to-one correspondence with the quantization group.  If needed, additional
constraints can be applied to the state vectors.

In what follows, we assume that the spacetime $\SpaceTime{M}$ can be locally
mapped homeomorphically onto a part of our structure group. The structure group is defined by its action in a complex Hilbert space
$\StateSpace{K}_0$ providing a continuous (in this paper unitary) representation
$T_{\Group{G}_0}\ni \hat{T}(g) \colon \StateSpace{K}_0 \to \StateSpace{K}_0$ for
$g \in \Group{G}_0$ in the carrier space $\StateSpace{K}_0$. This representation does not need to be, in general, irreducible. In the case
in which both groups (the quantization group and the structure group) coincide, this
representation has to be irreducible.

The quantization group $\Group{G}_Q \subset \Group{G}_0$ acts in a Hilbert space $\StateSpace{K} \subset \StateSpace{K}_0$, as an irreducible unitary representation
$T_{\Group{G}_Q} \ni \hat{T}(h) \colon \StateSpace{K} \to \StateSpace{K}$, where
$h \in \Group{G}_Q$. In other words, the operators $\hat{T}(h)$ of the
representation $T_{\Group{G}_Q}$ coincide with the operators $\hat{T}(g)$ restricted to
the quantization group---this is a required compatibility between realizations
of the structure and quantization groups.

In addition, the irreducible representation of the quantization group should be
generated from a single vector $\Ket{\bar{\Phi}_0} \in \StateSpace{K}$ called
the fiducial vector.  This vector is a free parameter of the IQ method.

We assume that the set of vectors $\Ket{h}=\hat{T}(h)\Ket{\bar{\Phi}_0}$,
where $h \in \Group{G}_Q$ generates a system of coherent states in Perelomov
sense \cite{Per}, see also \cite{Klauder88}.
The following operator admits the resolution of unity
\begin{equation}
\label{ResUnit}
\int_{G_Q}d\mu(h)\Ket{h}\Bra{h}=\UnitOp \,.
\end{equation}
This means that the representation $T_{\Group{G}_Q}$ allows for a construction of a set of
required coherent states and the corresponding resolution of unity in the state
space $\StateSpace{K}$.

As the second ingredient of the quantum model, we consider the manifold
$\SpaceTime{M}$ representing spacetime, endowed with the metric $\mathbf{g}$. Since geometric properties of a curved spacetime may vary from one point to another, we consider a local version of the integral
quantization, which in a way couples spacetime properties with the Hilbert state space
$\StateSpace{K}$. For this purpose, we construct a spacetime dependent family of fiducial vectors
$\Ket{\Phi_0(x)}$, where $x \in \SpaceTime{M}$.

If possible and required, one can also add to our description the second
subgroup $\Group{G}_M \subset \Group{G}_0$, called the manifold group, for which
one defines a mapping $\sigma_M \colon \SpaceTime{M} \to \Group{G}_M$, and which also relates,
in some way, the spacetime and the structure group $\Group{G}_0$. This completes all
required degrees of freedom of our quantum system.

The next step consists of constructing the coherent states in the carrier space
$\StateSpace{K}$ of the quantization group. For this purpose, let us decompose
each element $g$ of the structure group using the double coset structure
$\ConfSpace{X}_D := \Group{G}_Q \backslash \Group{G}_0 / \Group{G}_M$.

Every state $\Ket{\bar{\Phi}_0}$ defined as
\begin{equation}
\label{FidVectInt}
\Ket{\bar{\Phi}_0(x;s,q_M)}:=\hat{T}(sq_M)\Ket{\Phi_0(x)} \in
\StateSpace{K},
\end{equation}
which meets condition \eqref{ResUnit}, can be used as a fiducial
vector to build a set of coherent states with respect to the quantization group
$\Group{G}_Q$. The ``intrinsic'' fiducial vector $\Ket{\Phi_0(x)}$ contains
information about the spacetime $\SpaceTime{M}$, i.e.,
$\Ket{\Phi_0(x)} \colon \SpaceTime{M} \to \StateSpace{K}$.

For a given fiducial vector
$\Ket{\bar{\Phi}_0(x;s,q_M)}$, with
fixed $s \in \mathrm{rps}(\ConfSpace{X}_D)$, $q_M \in \Group{G}_M$ and
$x \in \SpaceTime{M}$, the vectors
\begin{equation}
\label{CohStates}
\Ket{x;hsq_M}:=\hat{T}(h)\Ket{\bar{\Phi}_0(x;s,q_M)} \,,
\end{equation}
where $h$ runs over the quantization group $\Group{G}_Q$, form a set of coherent
states in the state space $\StateSpace{K}$. Here $\mathrm{rps}$ is an abbreviation of ``representative'', i.e., $\mathrm{rps}(\ConfSpace{X}_D)$ denotes a set representatives of the double cosets.

The choice of the set $\mathrm{rps}(\ConfSpace{X}_D)$ depends on the physical model
under consideration. It is optimal, when the double cosets are labeled by
some properties of the physical system.
Every set of elements $\tilde{s}=\tilde{h}s\tilde{q}_M$,
where $\tilde{h} \in \Group{G}_Q$ and $\tilde{q}_M \in \Group{G}_M$, can serve as
a new set of representatives labeling double cosets $\ConfSpace{X}_D$.

We emphasize that construction \eqref{CohStates} admits a
local form of intrinsic fiducial vectors $\Ket{\Phi_0(x)} $, which are
parametrically dependent on the spacetime points $x$. On the other hand, if we assume, in addition, that $q_M=\sigma_M(x)$, the formalism allows us to connect the
arguments of the state functions $\psi \in \StateSpace{K}$ with the manifold
$\SpaceTime{M}$. One must remember that $\StateSpace{K}$ is the carrier
space of the representation of the quantization group.

Because the representation $T_{\Group{G}_Q}$ is irreducible, one has a
family of resolutions of the unit operator in the state space $\StateSpace{K}$,
\begin{equation}
\label{UnitRes}
\frac{1}{A_{\Phi(x;sq_M)}}
\int_{\Group{G}_Q} d\mu(h) \Ket{x;hsq_M}\Bra{x;hsq_M}=\UnitOp
\end{equation}
for every $s \in \mathrm{rps}(\ConfSpace{X}_D)$, $q_M \in \Group{G}_M$, and
$x \in \SpaceTime{M}$.

Local integral quantization of any ``local'' function $f(x;g)=f(x;hsq)$
on the structure group $\Group{G}_0$ at the spacetime point $x$ is given by a
``deformation'' of the resolution of unity \eqref{UnitRes} as follows
\begin{equation}
\label{LIQ}
f \to \hat{f}(x;sq_M) := \frac{1}{A_{\Phi(x;sq_M)}} \int_{\Group{G}_Q} d\mu(h)
\Ket{x;hsq_M} f(x;hsq_M) \Bra{x;hsq_M} \,.
\end{equation}
If needed, the generalization of \eqref{LIQ} to the positive operators measures
(POVM) can be found in \cite{paper1}.

It is useful to recall that the integral quantization method requires a mapping
from the classical configuration space $\ConfSpace{X}$ to the structure group
$\Group{G}_0$ and a mapping from the spacetime $\SpaceTime{M}$ to the set of the of
the fiducial vectors $\SpaceTime{M} \ni x \to \Ket{\Phi_0(x)}$. Using the group
action on these states, one obtains an infinite set of mappings
$\SpaceTime{M} \ni x \to \hat{T}(hs) \hat{T}(q_M)\Ket{\Phi_0(x)}$, where
potentially $q_M=\sigma_M(x)$. Otherwise, $q_M$ are arbitrary parameters. These
mappings correspond to different dynamical states of spacetime points. From this
point of view ``quantum points'' of the spacetime, represented by the coherent states, are degenerated in the state space.

Due to the above mappings, our semiclassical approach allows us to construct the
projection operators
$\hat{S}(x;hsq_M):=
\hat{T}(hsq_M)\Ket{\Phi_0(x)}\Bra{\Phi_0(x)}\hat{T}(hsq_M)^\dagger$. They
measure a ``localization'' of any quantum state $\Ket{\Psi}$ at the ``quantum
point'' $\hat{T}(hsq_M)\Ket{\Phi_0(x)}$ of the configuration space. In this
sense, the probability of finding the system at $x$ with the parameters
$h,s,q_M$ in the state $\hat{T}(hsq_M)\Ket{\Phi_0(x)}$ is given by
$\Bra{\Psi} \hat{S}(x;hsq_M) \Ket{\Psi}$.  However, one needs to notice that the
states $\hat{T}(hsq_M)\Ket{\Phi_0(x)}$ are not orthogonal, and they cannot be
considered as quantum alternatives. They do not allow for a construction of a
single operator representing position $x$ in spacetime. To obtain such an operator,
the quantization of spacetime is required and one has to use POVM quantum
measures.

It is important to realize that quantum observables, as well as classical observables,
are usually related to a chosen reference frame. At every reference frame, one can
construct the appropriate coordinate frame, required for the actual
application.

In our approach, the primary coordinate frame is defined by a choice of a
parametrization of the structure group. This parametrization remains ``agnostic'' about the
points of spacetime, and the corresponding coordinate frame is, in general, not a reference
frame. However, the mapping $\sigma_M$ introduces a set of reference points
$x \in \SpaceTime{M}$, which allow to fix the ``parametrization'' frame to points
of spacetime. The group composition law determines what modifications of the
coordinates are expected while changing variables describing group elements or
passing from a point $x$ to $x'$.

\section{Lorentz-Heisenberg-Weyl group} \label{LLL}

Having in mind the integral quantization based on the coherent states defined
by a representation of the Heisenberg-Weyl group $\HW{4}$, which remains close
to the canonical quantization, we extend the Heisenberg-Weyl group $\HW{4}$ by supplementing it with Lorentz transformations. In a sense, this addition renders the resulting group to be more ``relativistic''. For this purpose, we construct the
semidirect product of the Heisenberg-Weyl group and the Lorentz group.

As usual, the Heisenberg-Weyl group provides the phase space
$\mathcal{F}:= \{(p_\mu,q^\mu): \mu=0,1,2,3, (p_\mu,q^\mu) \in \RNumb^4 \times
\RNumb^4\}$ which, in this semiclassical approach, can be identified with a
cotangent bundle $T^* \SpaceTime{M}$ of the Schwarzschild spacetime
$(\SpaceTime{M}, \mathbf{g})$. A relation between this phase space and the
spacetime is defined by constructing the appropriate double cosets, as described in the
previous section.

The Heisenberg--Weyl group $\HW{4}$ is, in fact, a non-relativistic group of
transformations. In \cite{paper1,paper2} we used an external
action of the Lorentz group on the quantized observables. It acted on the
quantum numbers labeling the coherent states. Although some required properties were
satisfied, the full relativistic invariance needed a relativistically
invariant fiducial vector. This requirement remains in contradiction with its square
integrability in the carrier space $\StateSpace{K}=L^2(\RNumb^4,d^4w)$. To solve
this problem, we now extend the $\HW{4}$ group with the Lorentz group. We will denote
this group as $\LHW=\HW{4} \rtimes \Group{SO(3,1)}$.

The group $\LHW$ is constructed as a natural extension of the irreducible
representation of the group $\HW{4}$. It is defined as
\begin{equation}
\label{HWL0}
\hat{\mathcal{U}}(\kappa,p,q,\Lambda) =
\exp(i\kappa \UnitOp)\, \hat{\mathcal{U}}(p,q,\Lambda)\, ,
\end{equation}
where
\begin{equation}
\label{HWL1}
\hat{\mathcal{U}}(p,q,\Lambda) \psi(w)
=\exp\left(-i\frac{p\cdot q}{2} \right)
\exp\left( i\,p\cdot w \right)
\psi(\Lambda^{-1}[w-q]) \, .
\end{equation}
Here $p \cdot q:=p_\mu\,q^\mu$, $\UnitOp$ is the unit operator symbol, the
states $\Ket{\psi}$ are defined as
$\psi(w):= \BraKet{w}{\psi} \in \StateSpace{K}$, and $\Lambda{w}$ represents the
action of the Lorentz transformation on the four-vector $w$.

The group multiplication is given by
\begin{eqnarray}\label{HWL2}
&&\nonumber
\hat{\mathcal{U}}(\kappa',p',q',\Lambda') \hat{\mathcal{U}}(\kappa,p,q,\Lambda)
=\\
&&=
\hat{\mathcal{U}}\left(
\kappa'+\kappa -\frac{1}{2}
[(\Lambda'p)\cdot q'-p'\cdot(\Lambda'q)],\,
p'+\Lambda' p,\,
q'+ \Lambda' q,\,
\Lambda'\Lambda\right)
\end{eqnarray}
or, in terms of the elements $(\kappa,p,q,\Lambda)_G$ of the abstract version of the
group $\LHW$, by
\begin{eqnarray}
\label{HWL2a}
&&(\kappa',p',q',\Lambda')_G \, (\kappa,p,q,\Lambda)_G
= \\
&&=\left(\kappa'+\kappa -\frac{1}{2} [(\Lambda'p)\cdot q'-p'\cdot(\Lambda'q)],\;
p'+\Lambda' p,\,
q'+ \Lambda' q,\,
\Lambda'\Lambda\right)_G \,.
\end{eqnarray}
The neutral element of the group reads
\begin{equation}\label{HWL3}
\hat{\mathcal{U}}(0,0,0,\MatUnit)=\UnitOp \,,
\end{equation}
and the inverse elements are represented by
\begin{equation}\label{HWL4}
\hat{\mathcal{U}}(\kappa,p,q,\Lambda)^{-1}=
\hat{\mathcal{U}}\left(-\kappa,\,-\Lambda^{-1} p,\, -\Lambda^{-1} q,\,\Lambda^{-1}\right) \,.
\end{equation}
The action \eqref{HWL1} confirms that the group $\LHW$ is a semidirect product
of the Lorentz and the Heisenberg--Weyl groups, i.e.,
$\LHW=\HW{4}\rtimes\Group{SO(1,3)}$.
The Heisenberg-Weyl group is a normal subgroup of LHW, with the Lorentz group acting on its
phase-space parameters through the standard vector and dual vector representations.
Note that for $\Lambda=\MatUnit$, the operators
$\hat{\mathcal{U}}(\kappa,p,q,\MatUnit)$ defines an irreducible unitary representation of the Heisenberg--Weyl group in the Hilbert space
$\StateSpace{K}$.

Unitarity of \eqref{HWL1} can be checked directly, by calculating the
scalar product of the transformed states $\psi_1,\psi_2 \in \StateSpace{K}$,
\begin{eqnarray}
\label{RepUnitarity}
&& \BraKet{\hat{\mathcal{U}}(\kappa,p,q,\Lambda)\psi_2}%
          {\hat{\mathcal{U}}(\kappa,p,q,\Lambda)\psi_1}=
\int_{\RNumb^4} d^4{w} \nonumber\\
&& \left[e^{i\kappa} \exp\left(-\frac{i}{2}p \cdot q\right)
\exp\left(i\,p \cdot w\right)
\psi_2\left(\Lambda^{-1}(w-q)\right)\right]^\star \nonumber\\
&& \left[e^{i\kappa} \exp\left(-\frac{i}{2}p \cdot q\right)
\exp\left(i\,p \cdot w\right)
\psi_1\left(\Lambda^{-1}(w-q)\right)\right] =\BraKet{\psi_2}{\psi_1} \,,
\end{eqnarray}
where the invariance of the measure $d^4{w}$ with respect to translations and
Lorentz transformation is used. In the derivation \eqref{RepUnitarity} the
subsequent substitutions $w-q \to w$ and
$\Lambda^{-1}{w} \to w$ were applied.

The extension of the group $\HW{4}$ to the group $\LHW$ does not change the
irreducibility, because the carrier space $\StateSpace{K}$ of the
Heisenberg--Weyl group representations is the full space of square integrable
functions, and the extended action cannot be reducible. This feature generates a
family of equivalent $\HW{4}$ coherent states, which differ by the Lorentz
transformations of the fiducial vector
\begin{equation}
\label{HWLCS}
\BraKet{w}{\kappa,p,q,\Lambda}=\hat{\mathcal{U}}(\kappa,p,q,\Lambda) \Phi_0(w)=
e^{i\kappa}
\exp\left( -i\frac{p \cdot q}{2} \right)
\exp\left( i\,p \cdot w \right)
\Phi_0(x;\Lambda^{-1}[w-q]) \,.
\end{equation}
In this way, we get a family of resolutions of unity in $\StateSpace{K}$
labeled by $\Lambda$
\begin{equation}
\label{ResUnity}
 \frac{1}{(2\pi) ^4}
\int_{\RNumb^8}\, d^4p\, d^4q\, \Ket{\kappa,p,q,\Lambda}\Bra{\kappa,p,q,\Lambda} =\UnitOp \,.
\end{equation}
For every fixed $\Lambda$, the deformation of the measure generated by the
resolutions of unity \eqref{ResUnity} leads to integral quantization of any
classical real function determined on the classical phase space represented by
the group manifold of the Heisenberg--Weyl group (see \cite{paper1,paper2})
\begin{equation}
\label{IntQuantLamb}
f(p,q) \to \hat{f}_{\Lambda}:= \frac{1}{(2\pi ) ^4}
\int_{\RNumb^8}\, d^4p\, d^4q\,
\Ket{\kappa,p,q,\Lambda} f(p,q) \Bra{\kappa,p,q,\Lambda} \,.
\end{equation}
The mapping of the structure group onto the classical configuration space relates
functions on the structure group and their counterparts on the classical
configuration space: $f((\kappa,p,q,\Lambda)_G)\equiv f(p,q)$, for every
$\kappa$ and $\Lambda$. This implies that, choosing in \eqref{IntQuantLamb} the
group parameters $\Lambda=\MatUnit$, one obtains a canonical-type quantization,
generated by the Heisenberg--Weyl group, as in previous works \cite{paper1,paper2}.

It should be noted that in the above formulas \eqref{ResUnity} and
\eqref{IntQuantLamb}, the dependence on $\kappa$ is simplified.  For this
reason, the measure which ensures that coherent states satisfy the desired
properties was chosen as $d^4p\, d^4q$. For the same reason, we omit
the dependence on $\kappa$ in the notation for coherent states in the following
sections.

It is important to note that, instead of the operators \eqref{IntQuantLamb},
which often are merely symmetric, one can use the corresponding
selfadjoint POV measures, described in \cite{paper1}. In this case, all required
derivations will be very similar to those shown in that paper.

Transformation properties with respect to the Lorentz subgroup of the group $\LHW$
constitute important features of the operators \eqref{IntQuantLamb}. Let us denote the Lorentz transformation in our representation by
$\mathcal{U}(\tilde{\Lambda}) \equiv \mathcal{U}(0,0,0,\tilde{\Lambda})$. First,
we need the Lorentz transformation of the classical function $f(p,q)$. It is
defined in a standard way as
\begin{eqnarray}
\label{LTransFun}
&& \mathcal{U}(\tilde{\Lambda})f(p,q)
= \mathcal{U}(\tilde{\Lambda})f\left((\kappa,p,q,\Lambda)_G\right)
= f\left((0,0,0,\tilde{\Lambda})_G^{-1}(\kappa,p,q,\Lambda)_G\right) \nonumber\\
&& =f\left((\kappa,\tilde{\Lambda}^{-1}p,\tilde{\Lambda}^{-1}q,
\tilde{\Lambda}^{-1}\Lambda)_G\right)
=f\left(\tilde{\Lambda}^{-1}p,\tilde{\Lambda}^{-1}q\right) \,.
\end{eqnarray}
Next, we can calculate the Lorentz transformation of the
quantized operator \eqref{IntQuantLamb}:
\begin{eqnarray}
\label{LTransfOper}
&& \hat{\mathcal{U}}(\tilde{\Lambda})
\hat{f}_{\Lambda}
\hat{\mathcal{U}}(\tilde{\Lambda})^\dagger
=\frac{1}{(2\pi ) ^4} \int_{\RNumb^8}\, d^4p\,d^4q\,
\Ket{\tilde{\Lambda}p,\tilde{\Lambda}q,\tilde{\Lambda}\Lambda}
f(p,q)
\Bra{\tilde{\Lambda}p,\tilde{\Lambda}q,\tilde{\Lambda}\Lambda}
\nonumber\\
&& =
\frac{1}{(2\pi ) ^4} \int_{\RNumb^8}\, d^4p'\,d^4q'\,
\Ket{p',q',\tilde{\Lambda}\Lambda}
f'(p',q')
\Bra{p',q',\tilde{\Lambda}\Lambda} = \hat{f'}_{\tilde{\Lambda}\Lambda} \,,
\end{eqnarray}
where
\begin{equation}
\label{LHWActFun}
f'(p,q)=\hat{\mathcal{U}}(\tilde{\Lambda}) f(p,q)
= f\left(\tilde{\Lambda}^{-1}p,\tilde{\Lambda}^{-1}q\right)
\end{equation}
is the expected action of the Lorentz transformation onto functions on the phase
space. This action determines the covariance of our integral quantization with
respect to the Lorentz transformations.

\section{Model $\HW{4} \backslash \LHW/ \Group{T}(4)$}\label{Spec}

Let us consider the following triplet of groups: the structure group
$\Group{G}_0=\LHW$, the quantization group $\Group{G}_Q=\HW{4}$, and the
manifold group in the form of the spacetime translation subgroup
$\Group{T}(4) \subset \LHW$. The required configuration space will be constructed
as the double coset \cite{harmonic}.

The double cosets $\ConfSpace{X}_D=\HW{4} \backslash \LHW/ \Group{T}(4)$ can now be labeled by the Lorentz transformations $s=\Lambda$. Every element of the
group $\LHW$ can be decomposed as follows:
\begin{eqnarray}
\label{ElemCosetDecomp}
&&\hat{\mathcal{U}}(\kappa',p',q',\Lambda):=
\hat{\mathcal{U}}\left(\kappa +\frac{1}{2}p \cdot (\Lambda q_M),\,
p,\,
q+\Lambda q_M,\,
\Lambda\right) \nonumber\\
&& =\hat{\mathcal{U}}(\kappa,p,q,\MatUnit)
\hat{\mathcal{U}}(0,0,0,\Lambda)
\hat{\mathcal{U}}(0,0,q_M,\MatUnit) \,.
\end{eqnarray}
Using previous notation, the quantization group elements are
$\hat{T}(h)=\hat{\mathcal{U}}(\kappa,p,q,\MatUnit)$, the representatives of the
double coset are $\hat{T}(s)=\hat{\mathcal{U}}(0,0,0,\Lambda)$, and the manifold
group elements, denoted earlier by $\hat{T}(q_M)$, are now
$\hat{\mathcal{U}}(0,0,q_M,\MatUnit)$.

In this case, both Hilbert spaces coincide, i.e.,
$\StateSpace{K}_0=\StateSpace{K}$, and the representation of the structure group
$T_{\Group{G}_0}$ is also irreducible.

The coherent states labeled by $x$, $\Lambda$, $q_M$, $p$, and $q$ are generated by the
operators \eqref{ElemCosetDecomp}:
\begin{eqnarray}
\label{CSLHW}
&& \Ket{x;\Lambda,q_M;p,q}:=
\hat{\mathcal{U}}\left(\kappa +\frac{1}{2}p \cdot (\Lambda q_M),\,
p,\,
q+\Lambda q_M,\,
\Lambda\right) \Ket{\Phi_0(x)} \nonumber\\
&& =\hat{\mathcal{U}}(\kappa,p,q,\MatUnit)
\hat{\mathcal{U}}(0,0,0,\Lambda)
\hat{\mathcal{U}}(0,0,q_M,\MatUnit)
\Ket{\Phi_0(x)} \nonumber\\
&& =\hat{\mathcal{U}}(\kappa,p,q,\MatUnit)\Ket{\bar{\Phi}_0(x,\Lambda,q_M)} \,.
\end{eqnarray}
In the expression \eqref{CSLHW}, the parameters $x$, $\Lambda$, and $q_M$ label a family of
available coherent states.

In the fiducial vector $\Ket{\Phi_0(x)}$, we do not introduce any parametric
dependence on $\Lambda$ and $q_M$ to have covariant behaviour of the coherent
states with respect to the Lorentz transformations.  However, this fiducial
vector depends parametrically on spacetime points $x$. In this case, the action of
the Lorentz transformation $\hat{\mathcal{U}}(0,0,0,\Lambda')$ on the coherent
states \eqref{CSLHW} leads to a new coherent state
\begin{equation}
\label{LorentzCSLHW}
\hat{\mathcal{U}}(0,0,0,\Lambda')\Ket{x,\Lambda,q_M;p,q}
=\Ket{x,\Lambda'\Lambda,q_M;\Lambda' p,\Lambda' q}  \,.
\end{equation}
Formula \eqref{LorentzCSLHW} represents the local Lorentz transformation of
the state $\Ket{x,\Lambda,q_M;p,q}$ at the spacetime point $x$. Note the invariance
of $q_M$ with respect to the transformation
$\hat{\mathcal{U}}(0,0,0,\Lambda')$.

\subsection{Elementary observables in the  local integral quantization}

The generalized positions and the generalized momenta constitute standard elementary
observables for dynamical systems. In our model, the Heisenberg-Weyl subgroup
provides both observables.

In the carrier space $\StateSpace{K}=L^2(\RNumb^4,d^4w)$, the ``position''
four-vector operator is represented by the multiplication operators
$\hat{w}^\mu f(w):=w^\mu f(w)$, and the ``momentum'' four-vector operator
is built as the set of differential operators
$-i\hbar \partial_\mu \equiv -i\hbar \partial/\partial w^\mu$. The common
generalized eigenstates of the operators $\hat{w}^\mu$, denoted as
$\Ket{w}$, are four dimensional Dirac delta distributions
$\delta^4(w'-w)$. Similarly, the generalized eigenstates of the ``momentum''
four-vector operator $-i\hbar \partial_\mu$ are the functions
$\eta_k(w):=\frac{1}{(2\pi\hbar)^2} \exp(ik_\mu w^\mu/\hbar)$.

In our semi-classical approach to the spacetime $\SpaceTime{M}$, there are two
elementary quantum observables: the four-position $q^\mu$ and the dual variable
$p_\mu$ in the cotangent bundle $T^* \SpaceTime{M}$.

The quantization of  the four-vector $q$ at the point $x$ gives
\begin{equation}
\label{FourPosOp}
\hat{q}^\mu(x,\Lambda,q_M) =
\frac{1}{(2\pi ) ^4} \int_{\RNumb^8}\, d^4p\,d^4q\,
\Ket{x,\Lambda,q_M;p,q} q^\mu \Bra{x,\Lambda,q_M;p,q} \,.
\end{equation}
To clarify the structure of the above operator, let us calculate its matrix
elements within the states $\Ket{w}$. They are given by
\begin{eqnarray}
\label{ElMacFourPosOp}
&& \Bra{w''} \hat{q}^\mu(x,\Lambda,q_M) \Ket{w'} =
\delta^4(w''-w')\left( {w'}^\mu
- \int_{\RNumb^{4}} d^4q\, q^\mu |\bar{\Phi}_0(x,\Lambda,q_M;q)|^2 \right)
\nonumber\\
&& =\delta^4(w''-w')
\left({w'}^\mu -[(\Lambda q_M)^\mu + \AAver{x;(\Lambda q)^\mu ,\Phi_0}] \right) \, ,
\end{eqnarray}
where $\Phi_0(x;q)$ is the intrinsic fiducial vector, and
\begin{equation}
\label{AAver}
\AAver{x;f,\gamma}:=\int_{\RNumb^{4}} d^4q\, f(x,q) |\gamma(x,q)|^2
\end{equation}
is the expectation value of the function $f$ with respect to the probability
density $|\gamma(x,q)|^2$.

In the carrier space $\StateSpace{K} =L^2(\RNumb^4,d^4w)$, both fiducial vectors
$\bar{\Phi}_0$ and $\Phi_0$ are related by a simple formula
\begin{equation}
\label{PhiBarPhi}
\bar{\Phi}_0(x,\Lambda,q_M;w) = \Phi_0(x;\Lambda^{-1}[w-\Lambda q_M]) \,.
\end{equation}
Here, one can see that the parameters $q_M$ can relate the position of the fiducial
vector to a point in the spacetime. For example, this allows one to fix the maximum of the
fiducial vector at $x$.
Using the spectral theorem, the quantization of  the four-vector $q$ gives
\begin{equation}
\label{qSpectrDecomp1}
\hat{q}^\mu(x,\Lambda,q_M) =
\int_{\RNumb^4} d^4w
\Ket{w}\{ w^\mu-[(\Lambda q_M)^\mu
+\AAver{x;(\Lambda q)^\mu ,\Phi_0}] \}
\Bra{w} \,.
\end{equation}
The spectrum of the operator $\hat{q}^\mu(x,\Lambda,q_M) $, expressed in terms of
the quantum numbers $w^\mu$ is thus given by
$q^\mu=w^\mu-[(\Lambda q_M)^\mu + \AAver{x;(\Lambda q)^\mu ,\Phi_0}]$, where
$w^\mu \in \RNumb$.  The formula \eqref{qSpectrDecomp1}, expressed in terms of
positions $q^\mu$ in the cotangent bundle, can be rewritten as
\begin{equation}
\label{qSpectrDecomp2}
\hat{q}^\mu(x,\Lambda,q_M) =
\int_{\RNumb^4} d^4q
\Ket{q+\Lambda q_M + \AAver{x;\Lambda q,\Phi_0}}
 q^\mu
\Bra{q+\Lambda q_M + \AAver{x;\Lambda q ,\Phi_0}}  \,,
\end{equation}
where the corresponding eigenfunctions in the carrier space can be represented
as the Dirac $\delta$-functions
$\BraKet{w}{q+\Lambda q_M + \AAver{x;\Lambda q,\Phi_0}}= \delta^4(w-[q+\Lambda
q_M + \AAver{x;\Lambda q ,\Phi_0}])$.
The spectral decomposition  \eqref{qSpectrDecomp2} shows the dependence of these
eigenfunctions on spacetime points. This means that, for a fixed eigenvalue
$q=(q^0,q^1,q^2,q^3)$, the corresponding eigenfuction is spacetime dependent.

Symmetries of a given fiducial vector can dramatically reduce this dependence. For example, a good parity of the intrinsic fiducial vector with
with respect to every variable can remove the term
$\AAver{x;\Lambda q,\Phi_0} $. In this case, the matrix elements
\eqref{ElMacFourPosOp}, and so the operators $\hat{q}^\mu(x,\Lambda,q_M)$, can depend on $x \in \SpaceTime{M}$ only through $q_M=\sigma_M(x)$.

To be consistent with the action of the $\LHW$ group, the quantization of the
four-momentum requires its covariant components. In this case one gets
\begin{equation}
\label{FourMomOp}
\hat{p}_\mu(x,\Lambda,q_M) =
\frac{1}{(2\pi ) ^4} \int_{\RNumb^8}\, d^4p\,d^4q\,
\Ket{x,\Lambda,q_M;p,q} p_\mu \Bra{x,\Lambda,q_M;p,q} \,.
\end{equation}
Similarly, as above, the calculation of matrix elements of the momentum operator in
the momentum basis labeled by the wave vectors $k=(k_0,k_1,k_2,k_3)$ gives
\begin{eqnarray}
\label{ElMacFourMomOp}
&&\Bra{\eta_{k''}} \hat{p}_\mu(x,\Lambda,q_M) \Ket{\eta_{k'}} =
\delta^4(k''-k') \left( k'_\mu  - \int_{\RNumb^4} d^4k\, k_\mu
|\BraKet{\eta_k}{\bar{\Phi}_0(x,\Lambda,q_M)}|^2 \right) \nonumber\\
&& =\delta^4(k''-k') \left( k'_\mu - \AAver{x;(\Lambda k)_\mu, \tilde{\Phi}_0 }\right) \,,
\end{eqnarray}
where $\tilde{\Phi}(x)\equiv\BraKet{\eta_k}{\Phi_0(x)}$ is the Fourier transform
of the intrinsic fiducial vector $\Phi_0(x;w)$.  Similarly to the previous case, the
matrix elements \eqref{ElMacFourMomOp} allow us to write the spectral
decomposition of the four-momentum operator as
\begin{equation}
\label{pSpectrDecomp1}
\hat{p}_\mu(x,\Lambda) =
\int_{\RNumb^4} d^4k
\Ket{\eta_k}( k_\mu - \AAver{x;(\Lambda k)_\mu, \tilde{\Phi}_0 }) \Bra{\eta_k}  \,.
\end{equation}
The spectrum of the operator $\hat{p}_\mu(x,\Lambda) $, expressed in terms of the
quantum numbers $k_\mu$, is thus given by
$p_\mu=k_\mu-\AAver{x;(\Lambda k)_\mu, \tilde{\Phi}_0 }$, where
$k_\mu \in \RNumb$.  Formula \eqref{pSpectrDecomp1} expressed in terms of
four-momenta $p_\mu$ can be rewritten in the cotangent bundle as
\begin{equation}
\label{pSpectrDecomp2}
\hat{p}_\mu(x,\Lambda) =
\int_{\RNumb^4} d^4p
\Ket{\eta_{p+\AAver{x;(\Lambda k),\tilde{\Phi}_0 }}}
p_\mu
\Bra{\eta_{p+\AAver{x;(\Lambda k),\tilde{\Phi}_0 }}} \,.
\end{equation}
In the integral quantization the coherent states play a role of quantum points
in the quantum configuration space. It is important to check the meaning of
quantum numbers $p$ and $q$ labeling these states. For this purpose, one needs to
calculate expectation values of the operators $\hat{p}_\mu$ and $\hat{q}^\mu$ within these states. Using the matrix elements \eqref{ElMacFourMomOp} and
\eqref{ElMacFourPosOp}, direct calculations yield the following simple formulas
\begin{eqnarray}
&&\Bra{x,\Lambda,q_M;p,q} \hat{p}_\mu(x,\Lambda,q_M)
\Ket{x,\Lambda,q_M;p,q}=p_\mu \label{pCS}\, , \\
&&\Bra{x,\Lambda,q_M;p,q} \hat{q}^\mu(x,\Lambda,q_M)
\Ket{x,\Lambda,q_M;p,q}=q^\mu\, . \label{qCS}
\end{eqnarray}
These expectation values are independent of choice of the fiducial vector.
This means that quantum numbers $p$ and $q$ represent expectation values of the
four-momentum and four-position operators of a system in a coherent
state.

\section{Semiclassical motion of a particle}\label{Sem}

In this paper the spacetime positions and metric are not quantized and they are
considered to be classical functions. In reality, quantization can smear
positions in spacetime. Our main aim is to show that classical causal
geodesics described in previous sections follow the calculated quantum
probability distributions in spacetime.  For simplicity and for the sake of numerical accuracy, we present our tests only for radial
geodesics.

The configuration space of our system is represented by coherent states, which
are non-orthogonal to each other. This implies a kind of ``probability geometry''
determined by their overlaps and possibly an additional structure coming from the metric tensor.

As a tool for describing the motion of the test particle we choose quantum transition
amplitudes. The quantum probability amplitude between the initial state $\Psi_A$
and the final state $\Psi_B$ is defined by the scalar product
$\Amp(\Psi_A,\Psi_B):=\BraKet{\Psi_B}{\Psi_A}$. According to standard rules, the
amplitude of passing the system through a sequence of $N$ intermediate states
$(\phi^{(1)}_{k_{1}},\phi^{(2)}_{k_{2}},\dots,\phi^{(N)}_{k_{N}})$ is given by the
product of elementary amplitudes
\begin{eqnarray}
\label{SeqAmp}
&&\Amp(\Psi_A,\phi^{(1)}_{k_{1}},\phi^{(2)}_{k_{2}},\dots,\phi^{(N)}_{k_{N}},\Psi_B)
 :=\BraKet{\Psi_B}{\phi^{(N)}_{k_{N}}}
\Bra{\phi^{(N)}_{k_{N}}}
\dots
\Ket{\phi^{(2)}_{k_{2}}}
\BraKet{\phi^{(2)}_{k_{2}}}{\phi^{(1)}_{k_{1}}}
\BraKet{\phi^{(1)}_{k_{1}}}{\Psi_A}\nonumber\\
&& =\Bra{\Psi_B} P^{(N)}_{k_{N}} \dots
P^{(2)}_{k_{2}}P^{(1)}_{k_{1}} \Ket{\Psi_A} \,,
\end{eqnarray}
where $P^{(n)}_{k_{n}}:= \Ket{\phi^{(n)}_{k_{n}}}\Bra{\phi^{(n)}_{k_{n}}}$ are
projection operators.  The last line of \eqref{SeqAmp} allows one to denote this
amplitude as
$\Amp(\Psi_A, P^{(1)}_{k_{1}}, P^{(2)}_{k_{2}}, \dots, P^{(N)}_{k_{N}}, \Psi_B)$.

In general, because of a possible quantum interference at every step of the
evolution, these amplitudes have to be summed up over the intermediate allowed
states, giving the general formula
\begin{eqnarray}
\label{Amplitudes}
&& \Amp(\Psi_A,P^{(1)}, P^{(2)},\dots,P^{(N)},\Psi_B)
 =\sum_{k_{1},k_{2},\dots,,k_{N}}
\Amp(\Psi_A,\phi^{(1)}_{k_{1}},\phi^{(2)}_{k_{2}},\dots,
\phi^{(N)}_{k_{N}},\Psi_B) \nonumber\\
&& = \Bra{\Psi_B} P^{(N)} \dots P^{(2)}P^{(1)} \Ket{\Psi_A} \,,
\end{eqnarray}
where $P^{(n)}:= \sum_{k_n} P^{(n)}_{k_{n}}$ is an operator projecting onto the
subspace spanned by the intermediate states (quantum alternatives) at the
$n^{th}$ evolution step. Although in the derivation of \eqref{Amplitudes} the
intermediate states at a given evolution step must represent quantum
alternatives, i.e., a set of orthogonal states, the final formula is independent
of their choice. The only requirement is that they have to generate the same subspace
as the one spanned by the intermediate states at a given evolution step.
The expression \eqref{Amplitudes} suggests that after evolution step $N$, the
normalized state vector is given by
\begin{eqnarray}
\label{StateVecN}
\Ket{\Psi^{(N)}}
= \Norm{ P^{(N)}\dots P^{(2)}P^{(1)} \Ket{\Psi_A}}^{-1}
P^{(N)} \dots P^{(2)}P^{(1)} \Ket{\Psi_A} \,.
\end{eqnarray}
This construction follows the idea of the projection evolution presented in \cite{Gozdz2023,Gozdz2024}. In
this case, the projections $P^{(l)}$, where $l=1,2,\dots,N$, represent evolution
operators along the family of corresponding subspaces $P^{(l)}\StateSpace{K}$.

Our test particle is defined by the quantized geodesic Hamitonian
\eqref{gHam}
\begin{equation}
\label{QgHam}
\hat{H} = \frac{1}{2} g^{\mu\nu}(x^\alpha) \widehat{p_{\mu} p_{\nu}} \, .
\end{equation}
In the same way as for elementary observables, the  matrix elements of
$\hat{H}$ between states $\Ket{\eta_k}$ can be determined as
\begin{eqnarray}
&&\Bra{\eta_{k''}}\hat{H}\Ket{\eta_{k'}}=\delta^{4}(k'-k'') \nonumber \\
&&\frac{1}{2}g^{\mu\nu}(x)
\left[k'_\mu k'_\nu-k'_\mu\AAver{x;(\Lambda p)_\nu}-k'_\nu\AAver{x;(\Lambda p)_\mu}+
\AAver{x;(\Lambda p)_\mu(\Lambda p)_\nu}
\right] \, .
\end{eqnarray}
Using the spectral decomposition, one can obtain the following form of the quantum geodesic Hamiltonian:
\begin{eqnarray}
\label{QgHam2}
&&\hat{H} =\int_{\RNumb^4} d^4p\, \Ket{\eta_{p+\AAver{x;\Lambda p}}}
\frac{1}{2} g^{\mu\nu}(x) \left\{
p_{\mu}p_{\nu}  + \mathrm{Cov}(x;(\Lambda p)_\mu,(\Lambda p)_\nu) \right\}
\Bra{\eta_{p+\AAver{x;\Lambda p}}} \,,
\end{eqnarray}
where the covariance matrix is given by
\begin{equation}
\label{CovM}
\mathrm{Cov}(x;(\Lambda p)_\mu,(\Lambda p)_\nu)
= \AAver{x;(\Lambda p)_\mu(\Lambda p)_\nu}
-\AAver{x;(\Lambda p)_\mu} \AAver{x;(\Lambda p)_\nu} \,.
\end{equation}
Here, to shorten the notation, we have written $\AAver{x;f}$ instead of
$\AAver{x;f,\tilde{\Phi}_0}$. The physical meaning of variables used in the expectation
values informs which fiducial vector, either $\Phi_0$ or $\tilde{\Phi}_0$,
should be applied. The spectrum of the quantum geodesic Hamiltonian differs from
 classical values by the covariance matrix, which depends on the
fiducial vector used in derivations.

The classical constraint $H=-\frac{1}{2}m^2$ suggests that the actual quantum motion
can be described by eigenfunctions in the type of Klein-Gordon wave packets
$\Ket{\eta_{p+\AAver{x;\Lambda p}}}$, with
the eigenvalues $p_\mu$ satisfying the condition
\begin{equation}
\frac{1}{2}g^{\mu\nu}(x)\left(p_\mu p_\nu+\mathrm{Cov}(x;(\Lambda p)_\mu,(\Lambda p)_\nu)\right)=-\frac{1}{2}m^2 \, .
\end{equation}
However, in the quantum world, four-positions and four-momenta do not commute and,
due to the Heisenberg uncertainty relations, they are smeared. To
take this property into account, a more realistic description requires introducing
some generalization of the mass shell concept. We refer to this extension as the mass layer.

The mass layer for our test particle of zero spin and mass $m$ at the
point $x \in \SpaceTime{M}$ having the thickness $\epsilon$ is defined by the
spectrum of the geodesic Hamiltonian \eqref{QgHam2} as
\begin{equation}
\label{MassLayer}
\mathcal{J}_{m}(x) := \left\{ p\colon
g^{\mu\nu}(x)\left\{ p_{\mu} p_{\nu}
+\mathrm{Cov}\left(x;(\Lambda p)_\mu,(\Lambda p)_\nu\right) \right\}
\in [-(\bar{m}^2+\epsilon), -\bar{m}^2], p^0 \geq 0  \right\} \,.
\end{equation}
It is generated by smearing the spectrum of the quantum geodesic Hamiltonian
\eqref{QgHam2}.  The smearing is described by two positive parameters $m$ and
$\epsilon$. Taking the limit of $\epsilon \rightarrow 0$ leads to the commonly
used notion of the mass shell. This mass layer is compatible with the choice of
the metric signature $(-,+,+,+)$ and the orthochronous part of the Lorentz group
\cite{SW}. An additional practical feature of the finite thickness mass layer
is that it allows one to perform calculations in the Hilbert state space without
using distributions.

The mass layer implies that the projection operator defining possible states of
the particle of mass $ m\in\left[\bar{m},\;\sqrt{\bar{m}^2+\epsilon}\,\right] $, at the
spacetime point $x$, has to be constructed from the spectral measure of the
quantum geodesic Hamiltonian \eqref{QgHam2}. The corresponding operator at the
point $x$ can be written as
\begin{equation}
\label{ProjMassLayer}
\hat{P}(\mathcal{J}_{m}(x)) :=
\int_{\RNumb^4} d^4 k\, \Ket{\eta_{k+\AAver{x;\Lambda p}}}
\chi(k \in \mathcal{J}_{m}(x))
\Bra{\eta_{k+\AAver{x;\Lambda p}}}  \, ,
\end{equation}
where $\chi (k \in Q)= 1$ iff the relation $Q$ is satisfied and is equal to
$0$, otherwise. The projections $\hat{P}(\mathcal{J}_{m}(x))\Ket{\Psi_0}$, where
$\Ket{\Psi_0} \in \StateSpace{K}$, represent Klein--Gordon type wave packets
constructed from eigensolutions of the quantum geodesic Hamiltonian and can be
considered as the evolution of the initial state of the test particle with mass
$m$.

In the standard form of quantum mechanics, the best ``semiclasical'' states
representing particles with well-defined position and momentum are harmonic
oscillator coherent states, because they minimize the Heisenberg uncertainty
principle. This suggests the choice of the fiducial vector as the ground state of the four
dimensional harmonic oscillator. The simplest form of this ground state function
is a product of four Gauss functions, expressed in terms of quasi-Cartesian
coordinates (see Sec.~\ref{Cartesian}). Thus, the fiducial vectors can be written
as
\begin{eqnarray}
&&
\label{HO_Phi0w}
\Phi_0(x;w):= \prod_{\mu=0}^3
\left( \frac{2}{\pi \lambda_{\mu}(x,\Lambda)}\right)^{\frac{1}{4}}
\exp\left(-\frac{(w^\mu)^2}{\lambda_\mu(x,\Lambda)} \right)\, ,  \\
&&
\label{HO_Phi0p}
\tilde{\Phi}_0(x;p)=\BraKet{\eta_{p}}{\Phi_0(x)}
=\prod_{\mu=0}^3
\left( \frac{\lambda_{\mu}(x,\Lambda)}{2\pi} \right)^{\frac{1}{4}}
\exp\left(-\frac{\lambda_\mu(x,\Lambda)(p_\mu)^2}{4}\right) \,.
\end{eqnarray}
Note that the quasi-Cartesian coordinates paramatrize naturally the $\LHW$
group described in previous sections.

Expression \eqref{HO_Phi0p} is the Fourier transform of \eqref{HO_Phi0w}. The
only freedom in this fiducial vector is its width $\sigma$, related to $\lambda$
by the simple formula $\sigma^2=\lambda_\mu(x,\Lambda)$. In the following,
we fix the Lorentz transformations to $\Lambda=\MatUnit$, and we omit this symbol
in further notation, if possible. The diagonal Gaussian ensures that the expectation
values of $w^\mu$ and $p_\mu$ are equal to zero, i.e., $\AAver{x;w^\mu,\Phi_0}=0$
and $\AAver{x;p_\mu,\tilde{\Phi}_0}=0$. This simplifies all formulas and
renormalizes momenta to a more classical form (no additional $x$ dependent shift
is present). In particular, the covariance matrix is given by the simple diagonal
matrix
\begin{equation}
\label{CovMGauss}
\mathrm{Cov}(x;p_\mu,p_\nu)
=\delta_{\mu\nu} \frac{1}{\lambda_\mu(x)} \,,
\end{equation}
where $\Var{x;p_\mu}=\frac{1}{\lambda_\mu(x)}$ is equal to the variances of the four-momentum components.

In addition, due to our semiclassical approximation, to eliminate the shift
\eqref{CovMGauss} in the classical region of spacetime, we require that outside
the horizon the quantum term representing this shift vanishes, i.e.,
$g^{\mu\nu}(x)\mathrm{Cov}(x;(\Lambda p)_\mu,(\Lambda p)_\nu)=0$. This leads to
the following equation:
\begin{equation}
\label{gVar}
\sum_{\mu=0}^3 g^{\mu\mu}(x) \Var{x;p_\mu}=0  \,.
\end{equation}
The simplest possible solution, of this equation is given by
\begin{equation}
\label{gVar2}
\Var{x;p_\mu}=\frac{1}{\lambda_\mu(x)}=\frac{a(\mu)}{|g^{\mu\mu}(x)|}  \,,
\end{equation}
where $a(\mu)$ denote some constants (and not functions of $x$). These constants
satisfy the following condition
\begin{equation}
\label{gVar3}
\sum_{\mu=0}^3 \mathrm{sgn}(g^{\mu\mu}(x))|a(\mu)|=0 \,.
\end{equation}
The signum function is defined as: $\mathrm{sgn}(u)=1$, for $u\ge 0$ and
$\mathrm{sgn}(u)=-1$ for $u<0$.

A particular choice of the constants $a(\mu)$ fixes the fiducial vector uniquely. In our tests
of the model, we assume that $a(1)=a(2)=a(3)=a(0)/3$, which reproduces the case known from the previous analysis of the quantization in the
Minkowski spacetime (see \cite{paper1,paper2}). In quasi-Cartesian coordinates, see
Sec.\ \ref{Cartesian}, this assumption represents the  isotropy of the space and
leads to $g^{\mu\nu}(x)\mathrm{Cov}(x;p_\mu, p_\nu)=0$ above the black hole horizon.
Below the horizon, an additional non-zero term $g^{\mu\nu}(x)\mathrm{Cov}(x;p_\mu, p_\nu)\neq0$
is included in the mass layer (\ref{MassLayer}). The constant $a(0)$
is treated as a free parameter, which scales all widths of the fiducial vectors.

In the following, we consider only the simplest amplitude
$\BraKet{\Psi_2}{\Psi_1}$, where
\begin{itemize}
\item The initial state of the test particle is described by the coherent state
$\Ket{\Psi_0}=\Ket{x';\MatUnit,q'_M;p';q'}$ localized at $x'$ \,.
\item This state evolves according to the mass-layer constraint as a
Klein-Gordon type wave packet of the test particle. It is located at the same
point $x'$:
\begin{equation}
\label{MassLState}
\Ket{\Psi_1}=
\Norm{\hat{P}(\mathcal{J}_{m}(x'))\Ket{x';\MatUnit,q'_M;p';q'}}^{-1}
\hat{P}(\mathcal{J}_{m}(x'))\Ket{x';\MatUnit,q'_M;p';q'} \,.
\end{equation}
\item The final state is again given by the coherent state
$\Ket{\Psi_2}=\Ket{x'';\MatUnit,q''_M;p'';q''}$.
\end{itemize}
The resulting transition probability  is given by
\begin{eqnarray}
\label{TranProb}
&& |\mathcal{A}|^2=|\BraKet{\Psi_2}{\Psi_1}|^2
=\Norm{\hat{P}(\mathcal{J}_{m}(x'))\Ket{x';\MatUnit,q'_M;p';q'}}^{-2}
\left|
\int_{\RNumb^4} d^4k\, \chi(k \in \mathcal{J}_{m}(x')) \right. \nonumber\\
&& \left. \exp\left(i [(q''+q''_M)-(q'+q'_M)] k) \right)
\tilde{\Phi}_0(x'';k-p'')^\star \tilde{\Phi}_0(x';k-p')
\right|^2 \,,
\end{eqnarray}
where the norm of $\Ket{\Psi_1}$ is taken as the $L^2$--norm.
To compare our semiclassical motion with the classical one for radial timelike geodesics, we
apply the energy and angular momentum conservation laws to the initial and
final states. In these coherent states the energy and angular momentum are
conserved only in average.

To better visualize the probability distributions, we introduce approximate classical
expectation values of the particle position. Because the classical motion is
described by the position as a function of time, we have to define the conditional
probability of finding the particle at a required space position
$({x^1}'',{x^2}'',{x^3}'')$, under the condition that the particle at
a given final time $t''\equiv {x^0}''$ is located somewhere in space. This probability
is determined by the calculated conditional transition probabilities
\begin{equation}
\label{ProbAtTime}
\mathrm{prob}(x' \to x''|t'')
= \frac{|\BraKet{(t'',{x^1}'',{x^2}'',{x^3}'');q_M(x'');p'',q''}{x';q_M(x');p',q'}|^2}{
\sum_{({x^1}'',{x^2}'',{x^3}'')}
|\BraKet{(t'',{x^1}'',{x^2}'',{x^3}'');q_M(x'');p'',q''}{x';q_M(x');p',q'}|^2}
 \,,
\end{equation}
where $\Ket{x';q_M(x');p',q'}$ denotes the initial state of the particle, and
$\Ket{x'';q_M(x'');p'',q''}$, $x''=(t'',{x^1}'',{x^2}'',{x^3}'')$ denote its
possible final states at time $t''$.

Because of spherical symmetry of the Schwarzschild spacetime, for the radial motion we are only interested in the expectation values of the radial coordinate as a function of
time $\bar{r}(t)$. It can be defined as
\begin{equation}
\label{rAver}
\bar{r}(t''):= \sum_{({x^1}'',{x^2}'',{x^3}'')}
r'' \, \mathrm{prob}(x' \to x''|t'')
\,,
\end{equation}
where $r'' = \sqrt{({x^1}'')^2 +({x^2}'')^2 + ({x^3}'')^2}$.
In practice, we only sum over a
subset of grid points, providing approximate values of $\bar{r}(t)$. This approximation is sensitive to the size of the
grid. If the range of the grid is restricted too much, the obtained approximation of $\bar{r}(t)$ may not resemble the true shape of $\bar{r}(t)$.  In our figures, the approximated functions
$\bar{r}(t)$ are shown as colour dots, the light cones are plotted as black
lines, and the timelike geodesics are represented by red lines.

In our tests, we consider, when possible, the amplitudes for the particle moving initially
up and finally up, also up-to-down, and down-to-down where the the terms ``up'' and ``down'' refer to the directions pointing outward and inward with respect to the black hole. The most interesting are
spacetime regions close to the radial turning points and the horizon. The amplitudes
depend on many parameters: the Schwarzschild mass $M$, the mass and four-momenta of
the test particle, and $\bar{m}$ and $\epsilon$ which describe the mass
layer. The parameters $q_M$ are identified with the positions $x$ in spacetime. The variable $q$
canonically conjugated to the four-momenta is assumed to vanish, because in this work the spacetime is not quantized.  Due to our
semiclassical approach, we also fix the energy $E=-p_0$ and the angular momenta $l^2,l_z$
in the initial and final states. In reality, the Schwarzschild mass should be
much larger than the mass of the test particle. However, due to numerical
limitations, we can fulfill this condition only at some level.

In all figures, the basic parameters are chosen as follows: $M=1000$,
 $m^2=\bar{m}^2+0.4\epsilon$, $\bar{m}^2=2$, $E=0.8\,m$, $l^2=l_z=0$,
$\epsilon=0.01\bar{m}$, and $a(0)=0.1$.

\begin{figure}
  \includegraphics[scale=0.6]{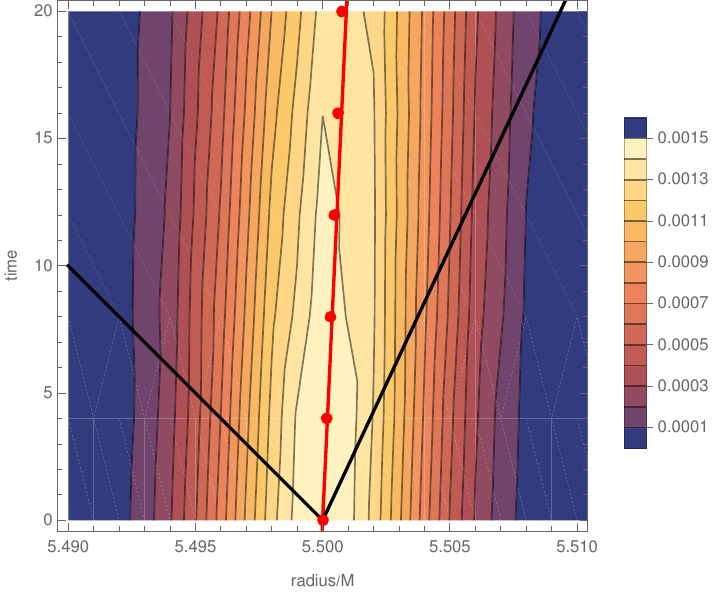}
  \includegraphics[scale=0.6]{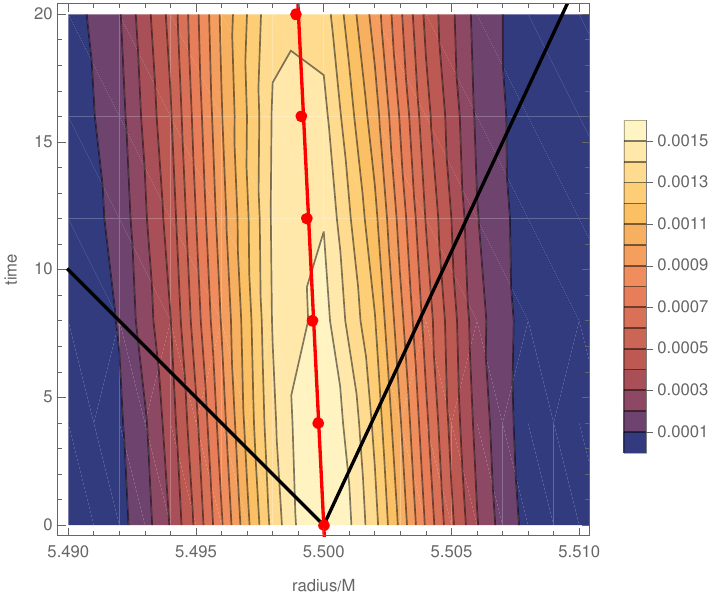}
  \caption{The probability distribution up-to-up (left panel) and down-to-down
  (right panel). On the left panel the ingoing geodesic starting from the
  initial point is plotted and on the right panel the corresponding outgoing
  geodesic is shown.}
\label{tr26Fig02}
\end{figure}

\begin{figure}
  \includegraphics[scale=0.55]{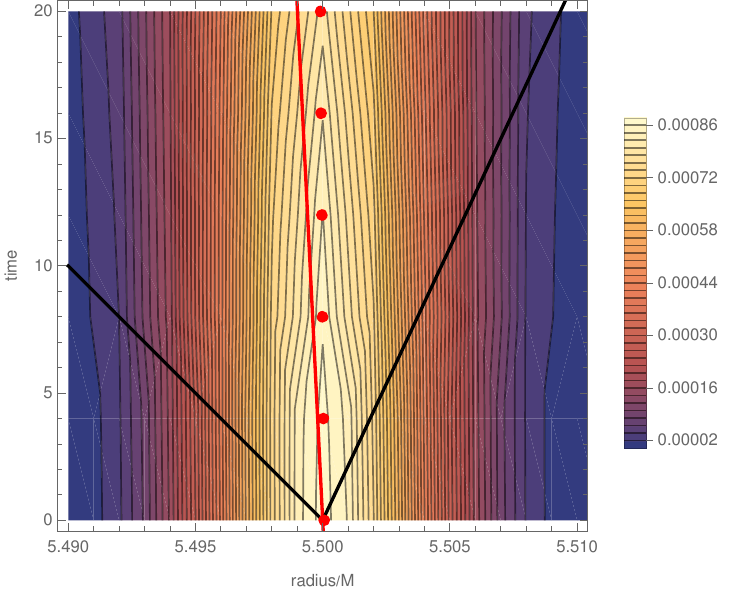}
  \includegraphics[scale=0.55]{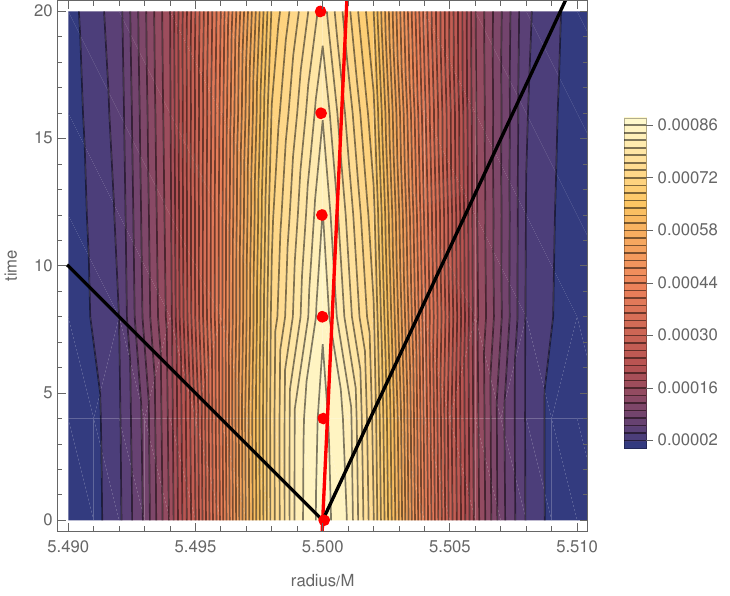}
  \caption{The probability distribution up-to-down. Comparison of the function
  $\bar{r}(t)$ to the ingoing geodesic (left panel) and to the outgoing
  geodesics (right panel). }
\label{tr27Fig02}
\end{figure}

In Figures~\ref{tr26Fig02} and \ref{tr27Fig02} the initial state is localized
at $\xi=\mathrm{radius}/M=5.5$ close to the turning point
$\xi_t=\mathrm{radius}/M=50/9$.
Figure~\ref{tr26Fig02} presents the probability distributions for the
amplitudes up-to-up (left panel) and down-to-down (right panel). The
average values of the radius follow the classical geodesics
Figure~\ref{tr27Fig02} shows a non-classical situation in which the initial
state corresponds to a particle is moving up, but at the same time we observe that with
some probability the particle is moving down.
This ``non-classic'' up-to-down amplitude, for which the initial state corresponds to the
motion up and final states to the motion down, describes a quantum effect caused
by the structure of both states, which are not eigenstates of the four-momentum
operators. They contain a mixture of both motions: up and down. The positions of the
expectation values $\bar{r}(t)$ differ significantly with respect classical geodesics, which suggests that, in this
case, there are non-zero probabilities to observe the competition of the two motions:
up-to-up and up-to-down.

\begin{figure}
 \includegraphics[scale=0.55]{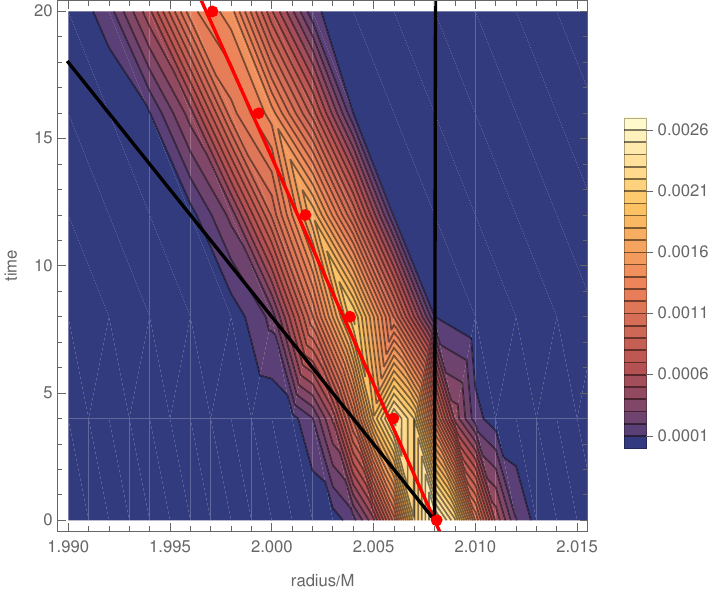}
 \includegraphics[scale=0.55]{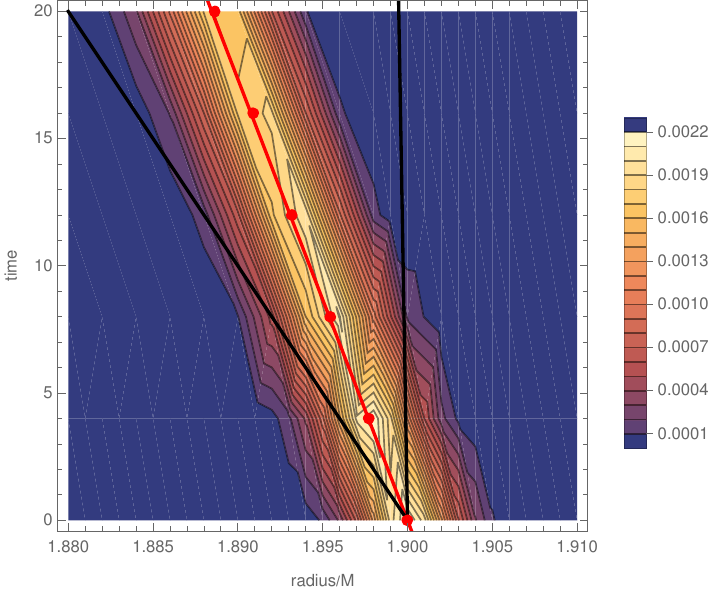}
  \caption{The probability distribution down-to-down starting from the point
  $\xi=\mathrm{radius}/M=2.008$ towards the points below the horizon (left
  panel). The probability distribution down-to-down starting from the point
  $\xi=\mathrm{radius}/M=1.9$ towards the singurality $\mathrm{r}=0$ (right
  panel) }
\label{tr10Fig01}
\end{figure}

As the second test of our semiclassical model, we compare its predictions with the geodesic motion of a particle passing through the horizon. The transitions
down-to-down are plotted in Fig.~\ref{tr10Fig01}. In the left panel the
initial state is localized just above the horizon at
$\xi=\mathrm{radius}/M=2.008$, and the particle is going down, plunging into the black hole.
%
%
As the last test we compared prediction of our model with the classical
results for the amplitudes down-to-down below the horizon (see the right panel in
Fig.~\ref{tr10Fig01}). In this case, the starting point is located at
$\xi = \mathrm{radius}/M = 1.9$

In all figures non-negligible probabilities can be found outside the light
cone. It is important to note that to reproduce classical results these
probabilities have to be taken into account, i.e., in all calculations we have
to include all amplitudes.

\section{Summary and Conclusions}\label{Con}

The main purpose of the present paper is to test the local integral quantization (LIQ)
and the related semi-classical model. As the test, we proposed a comparison between classical
radial causal geodesics and quantum probability amplitudes predicted by our model.
The proposed local integral quantization generalizes the
integral quantization used successfully earlier for the semiclassical description of
some aspects of gravity and also the quantum motion of particles in the Minkowski
spacetime. This generalization is tested and illustrated by its application to the
particle motion in the Schwarzschild spacetime.

In this paper the proposed local integral quantization is based on the group
$\Group{G}_0=\LHW$ referred to as the structure group, which is able to support both types
of degrees of freedom: classical variables describing the spacetime and quantum
variables describing quantum motion in this spacetime. As the structure group, we constructed an
extension of the Heisenberg-Weyl ($\HW{4}$) group, recovering some aspects of the
canonical quantization. This extension, denoted by $\LHW$, is
given by the semidirect product of the group $\HW{4}$ and the Lorentz group
$\Group{SO(3,1)}$, i.e., $\LHW=\HW{4} \rtimes \Group{SO(3,1)}$.

The quantum variables are related to a quantization group
$\Group{G}_{Q}=\HW{4} \subset \Group{G}_0$. The separation of variables into two
classes (classical and quantum variables) is done by a parametrization of
$\Group{G}_0$ by the double coset structure
$\DCosets{\HW{4}}{\LHW}{\Group{T}(4)}$, where the manifold group
$\Group{G}_M=\Group{T}(4) \subset \Group{G}_0$ is related to the spacetime. Such an
approach does not allow for the quantization of the spacetime itself, however, it
introduces a possibility to describe a semi-classical motion of particles in the
spacetime. Obviously, the same mathematical structure can be used with every
locally compact group, if it is required for a description of a physical
system. Possible extensions of the structure group to a non-compact case seem to be
also possible.

Local integral quantization of the variables $p,q$ related to the cotangent
bundle $\mathcal{T}^\star \mathcal{M}$ of our spacetime leads to quantum
operators with spectra differing from classical values by terms depending on
spacetime points. As it was shown, using some class of fiducial vectors, these
terms can be removed, if it is required by physical reasons. This adjustment can be compared to a
kind of renormalization of elementary variables to their classical counterparts.

Our test particles are spinless Klein-Gordon type objects, fully determined by
their masses, which in reality can be smeared. In this model, instead of a sharp
mass shell, we define and use a mass layer of a finite width. In a sense, this seems to be more physical
and allows one to omit the problem of operators of distributional values. From this point
of view, for a semi-classical description, the coherent states are the best
candidates for quantum states of the test particles. As natural evolution
operators, we consider operators $\hat{P}(\mathcal{J}_{m}(x))$, which project onto the
spacetime position dependent mass layer. Such projections lead to Klein-Gordon
type wave packets, evolving in the quantum configuration space.

In this paper we calculated only the simplest transition amplitudes between two coherent states, an initial and final, through an intermediate evolution step
generated by $\hat{P}(\mathcal{J}_{m}(x))$, which fixes the mass of the propagating
particle in a given mass layer and determines its Klein-Gordon type dynamics. It
turns out that the calculated probability distributions reproduce well the
classical motion of the test particle.





\end{document}